\documentclass[preprint,12pt, authoryear]{elsarticle}

\usepackage{aas_macros}
\usepackage{amssymb}
\usepackage{amsmath}
\usepackage{lineno}
\usepackage{calligra}

\usepackage{verbatim} 
\usepackage{csvsimple}
\usepackage{pgfplotstable}
\usepackage{multirow}
\usepackage{booktabs}
\usepackage{listings}
\usepackage{caption}
\usepackage{adjustbox}
\usepackage{subcaption}
\usepackage[colorlinks=true,linkcolor=blue]{hyperref}

\usepackage{algorithm}
\usepackage{algpseudocode}

\journal{Icarus}

\begin{document}

\begin{frontmatter}



\title{Orbital Stability of Closely-Spaced \\Four-planet Systems}

\affiliation[label1]{organization={South Dakota School of Mines and Technology},
            addressline={501 E St. Joseph St},
            city={Rapid City},
            postcode={57701},
            state={SD},
            country={USA}}

\affiliation[label2]{organization={Stevens High School},
            addressline={4215 Raider Road},
            city={Rapid City},
            postcode={57702},
            state={SD},
            country={USA}}
            
\affiliation[label3]{organization={Space Science \& Astrobiology Division, NASA Ames Research Center},
            addressline={MS 245-3},
            city={Moffett Field},
            postcode={94035},
            state={CA},
            country={USA}}

\author[label1]{Bennet Outland}
\author[label1]{Gretchen Noble}
\author[label2]{Andrew W. Smith}
\author[label3]{Jack J. Lissauer}

 
 


\begin{abstract}
We investigate the orbital dynamics of four-planet systems consisting of Earth-mass planets on initially-circular, coplanar orbits around a star of one solar mass. In our simulations, the innermost planet's semimajor axis is set at 1 AU, with subsequent semimajor axes spaced equally in terms of planets' mutual Hill radii.  Several sets of initial planetary longitudes are investigated, with integrations continuing for up to $10^{10}$ orbits of the innermost planet, stopping if a pair of planets pass within 0.01 AU of each other or if a planet is ejected from the system.  We find that the simulated lifetimes of four-planet systems follow the general trend of increasing exponentially with planetary spacing, as seen by previous studies of closely-spaced planets. Four-planet system lifetimes are intermediate between those of three- and five-planet systems and more similar to the latter. Moreover, as with five-planet systems, but in marked contrast to the three-planet case, no initial semimajor axes spacings are found to yield systems that survive several orders of magnitude longer than other similar spacings. First- and second-order mean-motion resonances (MMRs) between planets correlate with reductions in system lifetimes. Additionally, we find that third-order MMRs between planets on neighboring orbits also have a substantial, though smaller, destabilizing effect on systems very near resonance that otherwise would be very long-lived. Local extrema of system lifetimes as a function of planetary spacing occur at slightly smaller initial orbital separation for systems with planets initially at conjunction relative to those in which the planets begin on widely-separated longitudes. This shift is produced by the asymmetric mutual planetary perturbations as the planets separate in longitude from the initial aligned configuration that cause orbits to spread out in semimajor axis.
\end{abstract}




\end{frontmatter}


\section{Introduction} \label{sec:intro}

Over the past two decades, $\sim 1000$ multi-planet systems have been discovered, most by NASA's {\it Kepler} and {\it TESS} spacecraft and radial velocity observations. Analysis of photometric data from {\it Kepler}'s prime mission has revealed 709 candidate multi-planet systems, many of which include planets on closely-spaced orbits \citep{2024PSJ.....5..152L}. \textcolor{black}{Why multi-planetary systems of this nature form has been explored through planet-planet scattering models by \cite{raymond2009planet} and \cite{raymond2010planet}. The stability impacts of planet-planet scattering are investigated by \cite{marzari2025planet}. In \cite{raymond2008observable} in situ accretion, type 1 migration, gas giant shepherding, and secular resonance shepherding formation models were investigated for close-in planets. Moreover, an analysis of the formation of compact giant planets was completed by \cite{marzari2002eccentric}.} Pioneering numerical calculations of the stability of multi-planet systems were done at the dawn of the exoplanet era by \cite{1996Icar..119..261C}, followed by numerous subsequent studies, including several that focused on systems of three or five Earth-mass planets orbiting a solar mass star, such as \cite{SL09}, \cite{Obertas},  and \cite{2021MNRAS.506.6181B}. \cite{Smith_Lissauer_2010} studied systems with many co-orbital planets. However, the stability of four-planet systems has not been studied extensively, although \cite{rice2023stable} considered four-planet systems with non-equal masses. We numerically investigate the orbital dynamics of tens of thousands of hypothetical four-planet systems consisting of Earth-mass planets on initially-circular, geometrically-spaced, coplanar orbits around a star of one solar mass over a wide range of initial semimajor axes spacings and with several different initial longitudinal prescriptions. This is done to also study the effects of initial longitudes on the system lifetimes and thereby build upon the investigation of \cite{lissauer_gavino_2021}. 

In Section \ref{sec:methods}, we describe the integration techniques used and the orbital parameter space investigated. Results of our four-planet integrations are presented in Section \ref{sec:results}, together with parameters for exponential fits to system lifetime as a function of initial orbital separation for various sets of initial longitudes and number of planets in the system, where the exponential trend was analytically elucidated in \cite{petit2020path}. Section \ref{sec:345} compares lifetimes of systems with three, four and five planets and examines the effects of mean-motion resonances (MMRs). When comparing the lifetimes of systems with different initial longitudinal spacings, we observe a ``phase-shifting'', an offset in the orbital spacing corresponding with local maxima or minima in lifetimes, explored in Section \ref{phase_shifts}. A summary of our results and their implications can be seen in Section \ref{sec: conclusion}.

\section{Methods} \label{sec:methods}

Each of the simulated planetary systems consists of a star with one solar mass, M\textsubscript{$\bigodot$}, that is orbited by (in most cases four) planets, each with mass equal to that of Earth, M\textsubscript{$\bigoplus$}. The planets are coplanar and travel in the same direction on initially-circular orbits. The innermost planet has an initial semimajor axis of 1 AU, and subsequent planets are spaced such that the initial orbital period ratios of neighboring planets are equal, in keeping with several previously published studies such as \cite{SL09}, \cite{Obertas}, \cite{lissauer_gavino_2021}, \cite{2021MNRAS.506.6181B}, and \cite{GRATIA2021114038}. 

Numerical integrations in this work assume Newtonian mechanics and are performed using a symplectic (Hamiltonian-based) integrator. The systems are integrated step-wise using an implementation of the Wisdom-Holman algorithmic integrator, WHFast, in the Rebound integration package by \cite{reboundwhfast}. \textcolor{black}{A simulation time step of $1/(20 + 1/\lambda) \approx 0.0485$ times the inner planet's initial orbital period was used}, where $\lambda$ is the golden ratio:

\vspace{-2.5mm}

\begin{equation}
    \lambda \equiv \frac{5 ^ {\frac{1}{2}} + 1}{2} .
     \label{eq:lambda}
\end{equation}

\noindent Thus, the time step is slightly less than 5\% of the inner planets' orbital period, with the ratio being an irrational number to minimize the possibility of a resonance between the number of orbits and the number of time steps. This time step further strikes a balance between computational efficiency and accuracy, following \cite{GRATIA2021114038}.

\subsection{Initial Semimajor Axes} \label{subsec:initial Semimajor axes}
 
We measure the initial semimajor axis separation of neighboring planets in units of their mutual Hill radius, $R_{H_{j,j+1}}$, using the following recursive equation to calculate the mutual Hill radius for each neighboring pair of planets: 

\vspace{-2.5mm}

\begin{equation}
    R_{H_{j,j+1}} = \left[\frac{2M\textsubscript{$\bigoplus$}}{3M_{j}} \right] ^ {\frac{1}{3}} \frac{(a_{j} + a_{j + 1})}{2} = \left[\frac{2M\textsubscript{$\bigoplus$}}{3(M\textsubscript{$\bigodot$} + (j-1) M\textsubscript{$\bigoplus$})} \right] ^ {\frac{1}{3}} \frac{(a_{j} + a_{j + 1})}{2},
    \label{eq:Hill_radius}
\end{equation}

\noindent where $a_{j}$ is the semimajor axis of the $j$\textsuperscript{th} planet. The innermost planet has $j=1$ and the index of each succeeding planet is one greater than that of its immediate inner neighbor. The mass within the orbit of the $j$\textsuperscript{th} planet is denoted $M_{j}$ and is given by:

\vspace{-2.5mm}

\begin{equation}
    M_{j} = M_{\star} + \sum_{k=1} ^{j-1} m_{k} = M\textsubscript{$\bigodot$} + (j-1) M\textsubscript{$\bigoplus$},
\end{equation}

\noindent where $M_{\star}$ and $m_k$ are the central star's mass and the mass of the $k$\textsuperscript{th} planet (whose orbit is interior to that of the $j$\textsuperscript{th} planet), respectively. We define a non-dimensional quantity, $\beta$, to specify the spacing between the initial semimajor axes of neighboring planets:

\vspace{-2.5mm}

\begin{equation} \label{eq:beta_def}
    \beta \equiv \frac{a_{j+1} - a_{j}}{R_{H_{j,j+1}}}.
\end{equation}

\noindent Substituting Equation (\ref{eq:Hill_radius}) into Equation (\ref{eq:beta_def}) yields a recursive equation for the semimajor axes of the planets:

\begin{equation}
    a_{j} + \beta \left(\frac{M\textsubscript{$\bigoplus$}}{12 (M\textsubscript{$\bigodot$} + (j-1) M\textsubscript{$\bigoplus$})}\right)^{\frac{1}{3}} \frac{(a_{j} + a_{j + 1})}{2} = a_{j+1}.
    \label{eq:recursive}
\end{equation}

\noindent Algebraic manipulations of Equation (\ref{eq:recursive}) yield:

\begin{equation}
    a_{j+1} = a_{j} \left[1+\beta \left(\frac{M\textsubscript{$\bigoplus$}}{12 (M\textsubscript{$\bigodot$} + (j-1) M\textsubscript{$\bigoplus$})}\right) ^ {\frac{1}{3}} \right] \left[1 - \beta \left(\frac{M\textsubscript{$\bigoplus$}}{12 (M\textsubscript{$\bigodot$} + (j-1) M\textsubscript{$\bigoplus$})}\right) ^ {\frac{1}{3}} \right] ^{-1},
    \label{eq:semi_major_axis}
\end{equation}

\noindent which can be used to calculate the initial semimajor axes of the $(j+1)^{\rm th}$ planet from that of its inner neighbor. 

\subsection{Initial Longitudes} \label{subsec:initial longitudes}
\begin{figure}[t!]
    \begin{center}
    \includegraphics[width=100mm,scale=0.75]{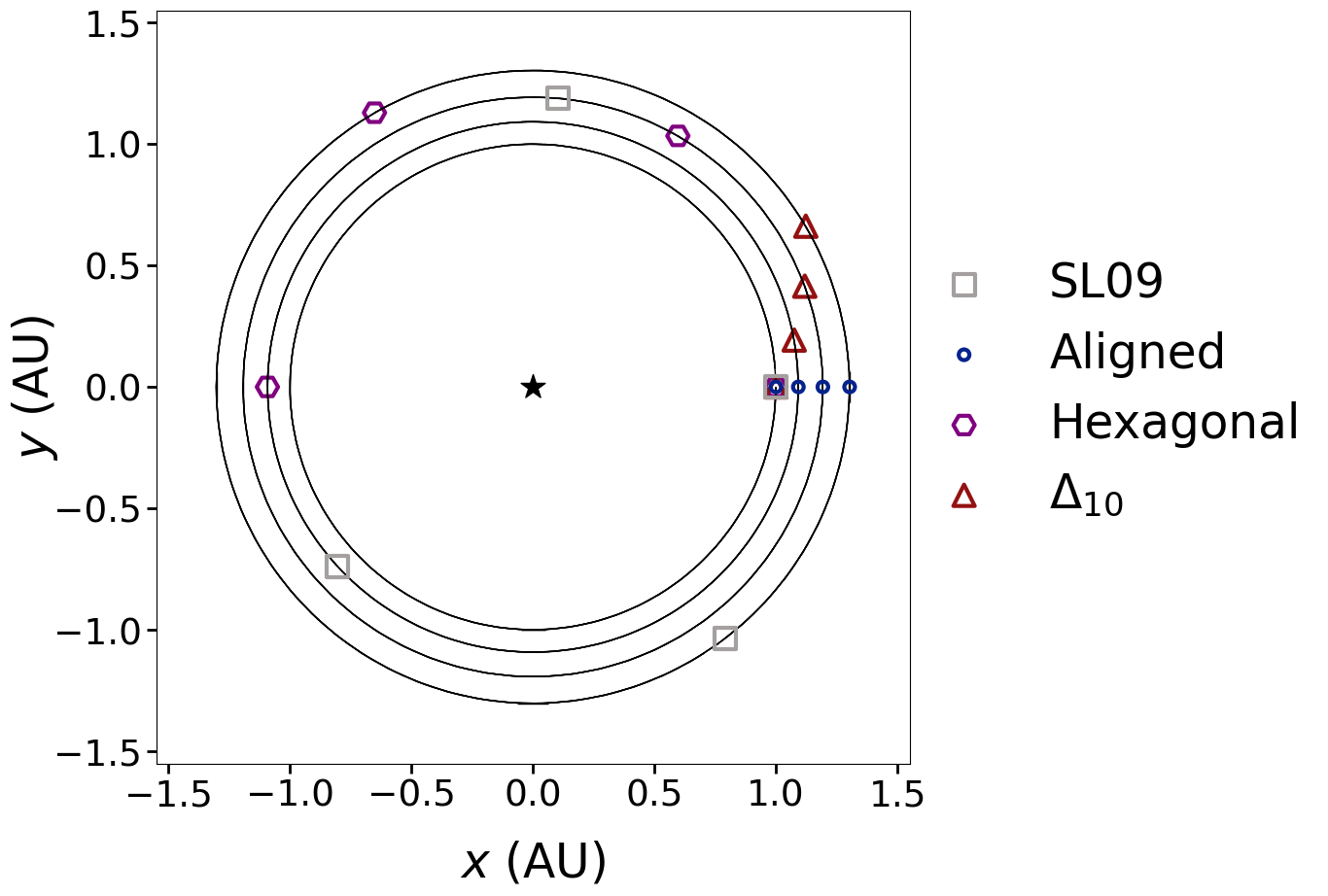}
    \end{center}
    \caption{Initial angular locations for four longitude prescriptions used at $\beta$ = 7.000, corresponding to a period ratio of $P_{i+1} / P_{i} \approx 1.142$. The five-pointed star at the center represents a one solar mass star, the rings trace the initially-circular planetary orbits, and the colored shapes show the initial planetary positions for the four initial sets of planetary longitudes listed in the legend on the right.  All planets orbit in the counterclockwise direction. 
    \label{fig:longitudes}}
\end{figure}

Given the assumptions of coplanar, initially-circular orbits, the only remaining orbital parameters that we need to specify are the initial orbital longitudes for each planet. We choose coordinates so that this angle is zero for the innermost planet. For the remaining planets, we investigate multiple sets of initial longitudes, including the longitudes used by \cite{SL09} (henceforth referred to as the SL09 longitudes), Aligned longitudes (all planets with a true anomaly of zero), Hexagonal longitudes, three distinct sets of longitudes with the planets approaching conjunction, and random, independently-drawn, longitudes. The formulas for each set of longitudes are given in Table \ref{tab:longitudes_table}, and a visual representation of four of these longitude combinations (including the most widely-spaced set of longitudes approaching conjunction, $\Delta_{10}$) is displayed in Figure \ref{fig:longitudes}. 

\begin{table}[h!]
    \centering
    \caption{Initial longitude prescriptions (in radians) used for our investigation into four-planet systems. For each of the formulas, $j$ is the planetary index. }
    \begin{tabular}{c|c}
        Longitudes & Formula \\
        \hline
        SL09 & $\theta_{SL} = 2\pi\lambda (j-1)$ \\ 
        Aligned & $\theta_{A} = 0$ \\
        Hexagonal & $\theta_{H} = \frac{\pi}{3}(j-1)$ \\
        $\Delta_1$ & $\theta_{\Delta, 1} = \frac{1}{5}\pi\lambda(j-1) (\frac{2 \pi}{360})$ \\ 
        $\Delta_5$ & $\theta_{\Delta, 5} = \pi\lambda(j-1) (\frac{2 \pi}{360})$ \\
        $\Delta_{10}$ & $\theta_{\Delta, 10} = 2\pi\lambda(j-1) (\frac{2 \pi}{360})$ \\
        Random & $\theta_R =2\pi\mathcal{U}(0,1)$ for $j\neq1$\\
    \end{tabular}
    \label{tab:longitudes_table}
\end{table}

SL09 and Aligned longitudes are selected to study planetary systems with disparate initial angular positions. Whereas SL09 longitudes are well-separated in terms of the true anomalies, the opposite is the case with the Aligned longitudes, where the planets share the same true anomaly. Such choices in longitudes are used to represent the extremes in initial interactions between planets. 

The Hexagonal longitudes provide an additional set of well-spaced longitudes and have planets placed at angles corresponding to the vertices of a regular hexagon. Following the procedure of \cite{lissauer_gavino_2021}, the inner three planets had longitudes of $0$, $\pi$, and $\frac{\pi}{3}$. For the  longitude of the outer planet, we selected $\frac{2\pi}{3}$, so that each pair of neighboring planets has a different separation in longitude and no pair of planets begins at conjunction. 

We find that well-spaced initial longitudes, such as SL09 and Hexagonal, yield planetary systems with very similar lifetime trends in terms of stability, but larger differences are found between well-spaced and Aligned systems. To explore whether the Aligned systems represented a special case, three additional sets of nearly aligned initial longitudes were studied.  The first set places each neighboring pair of planets approximately $10^\circ$ prior to conjunction, which is the same algorithm as for the ``Primary'' longitudes used for three-planet systems by \cite{lissauer_gavino_2021}. The remaining sets have longitudes computed by dividing the aforementioned initial longitudinal separations by two and ten. For the sake of brevity, these longitudes will be referred to as $\Delta_{10}$, $\Delta_5$, and $\Delta_{1}$, respectively (so the Aligned simulations if expressed using this notation would be $\Delta_{0}$). The full values of these longitudes given in Table \ref{tab:longitudes_table}, are rounded to the $16^{\mathrm{th}}$ decimal place for integration. 

Finally, we simulate systems with longitudes determined by pseudo-randomly assigning a value between 0 inclusive and $2 \pi$ exclusive, with each value being equally probable and chosen independently. Such values are redrawn for each planet and each mutual Hill radius increment. This is accomplished utilizing a pseudo-random uniform distribution and scaling the result by a factor of $2\pi$. 

\subsection{Integrator and End Conditions} \label{subsec:Integrator and End Conditions}
The symplectic Wisdom-Holman integrator, WHfast by \cite{reboundwhfast}, is used to perform integrations within the Rebound package from \cite{rebound}. Integrations are stopped and the lifetime of the system recorded once any one of three conditions are met:

\begin{enumerate}
    \item The Euclidean distance between two planets becomes less than or equal to 0.01 AU, which we define as a close encounter;
    \item A planet is ejected from the system, when it exceeds a pre-determined distance away from the star;
    \item The simulated lifetime of the system reaches 10\textsuperscript{10} years (initial orbital periods of the innermost planet).
\end{enumerate}

\textcolor{black}{Since two planets passing within 0.01 AU, which is approximately the mutual Hill sphere of two planets each of one Earth mass, of each other would cause a large orbital perturbation and rapid subsequent orbital stability degradation, a close encounter condition was selected as one termination condition.} \textcolor{black}{The presence of destabilizing close encounters are expected after formation as indicated by \cite{chatterjee2008dynamical}.}  Furthermore, there is also the possibility that the planets would collide, and the WHFast integrator neglects the full effects of very close bodies due to the underlying assumptions of the model as investigated by \cite{wh}. 

Another end condition that we considered using was an orbital "crossing" occurring in the system. This is accomplished by comparing the semimajor and semiminor axes of neighboring planets. For a given pair of neighboring planets, if the semimajor axis of the inner planet is greater than or equal to the semiminor axis of the outer planet, the orbits have "crossed" and are considered to be unstable. We validated the close encounter method against the orbital crossing method and found that the two approaches yielded very similar results. However, the orbit crossing method is significantly slower than the close encounter method, since additional data must be analyzed from the simulation before further progress can resume. While this load can be reduced by analyzing the orbital data less frequently, doing so would reduce the accuracy of the results.  

Ultimately, the close encounter condition was selected owing to improved simulation speed with nearly identical results. A drawback of this method is that it can fail to detect when the outer planet becomes ejected from the system with minimal perturbations to the other planets. While rare, this anomaly occurred at a mutual Hill radius of 3.493 in the four-planet SL09 dataset and caused a spuriously long-lived system, due to the residual three-planet system being considerably more stable than its four-planet counterpart as will be discussed in Section \ref{sec:exponential fits comparison}. To combat this, a planetary escape boundary was applied at a distance of double of the initial semimajor axis of the planet farthest from the star. However, the vast majority of systems did not terminate due to planetary ejection. \textcolor{black}{Planetary ejections of outer planets has been observed in \cite{rasio1996dynamical}}.

Finally, a lifetime limit was used to conserve computational resources. With the hardware available, approximately two weeks of computational time are required to reach a simulated lifetime of $10^{10}$ years in a four-planet system. Moving past this point for termination would not have been feasible with our computing resources; extending to $10^{11}$ years would take up to 20 weeks for a process to finish a single simulation. Despite this limitation, we were still able to demonstrate trends in simulated lifetimes as well as structure related to MMRs for a considerable range of initial orbital separation.

\subsection{Shifted Linear Regression} \label{sec:exponential fits comparison}

Once the datasets of the simulated lifetimes are obtained, we compare the stability of the different systems. Given that the relationship between lifetimes and initial orbital separation is well-approximated by an exponential, we take the (base ten) logarithm of the lifetimes to calculate a linear regression of the data. This equation can be modeled as such:

\vspace{-2.5mm}
\begin{equation}
    \textcolor{black}{\log t_{\rm c} = b\beta + c.}
\end{equation}

\noindent As only systems with $\beta > 2\sqrt{3}$ are considered, values of the constants $b$ and $c$ are highly correlated. To reduce this correlation, we follow \cite{QuarlesLissauer2018} by subtracting the two-planet orbital stability limit of $2\sqrt{3}$ from the mutual Hill radii spacing, effectively defining a new variable $\beta' \equiv \beta-2\sqrt{3}$ for our exponential fits. These fit parameters are represented following:

\vspace{-2.5mm}
\begin{equation}
    \textcolor{black}{\log t_{\text{c}} = b'(\beta-2\sqrt{3}) + c'.}
\end{equation}

\noindent These shifted coefficients, $b'$ and $c'$, are used for the comparison between the different longitudes and with differing numbers of planets. For this work, we utilized an ordinary least squares approach to determine the coefficients of the linear regression.

\subsection{Dispersion Metrics} \label{sec: Dispersion Metrics}
Partially utilizing the linear regression defined in the previous section, two metrics are created to measure the chaotic scatter inherent to the systems. We wish to measure such scatter through two methods: first, global scatter and second, local scatter. Following \cite{lissauer_gavino_2021}, a metric for the global scatter is determined via the standard deviation from the trend line:

\vspace{-2.5mm}
\begin{equation}
    \sigma_{\rm exp} = \text{std}\left( \{ \log t_{\rm c}(\beta) - b' \beta + c' \,\, \forall \beta \in B \}  \right),
\end{equation}

\noindent where $t_c(\beta)$ is the system lifetime as a function of the initial separation, $B$ is the set of initial separations in units of mutual Hill radii, and std is the standard deviation. This metric is the standard deviation of the difference between the log-lifetimes and the linear regression of the log-lifetimes for all initial separations considered. For the local metric, the distance from the rolling median is given by

\vspace{-2.5mm}
\begin{equation}
    \sigma_{\rm local} = \text{std}\left( \{ \log t_{\rm c}(\beta) - \text{RollingMedian}(t_{\rm c}(\beta), \beta, K) \,\, \forall \beta \in B \}  \right),
\end{equation}

\noindent where $K$ is the size of the 1D kernel. In this study, we define local to be a closed interval of width 0.02 mutual Hill radii centered upon the system in question. This implies  a kernel of $K=21$ points if the resolution in the initial spacing is 0.001 mutual Hill radii, as is the case in many of our studied systems (see \S \ref{sec:results}). If the resolution is 0.0005, then a kernel of $K=41$ points is used.  
Note that a resolution of 0.01 is not sufficient because the kernel would consist of only $K=3$ points, so we do not calculate $\sigma_{\rm local}$ for our low-resolution sets of runs. \textcolor{black}{This metric is analogous to the method used in \cite{10.1093/mnras/sty2418} to assess the distribution of lifetimes. However, the method presented also works for non-randomized datasets.}

The rolling median is used as a smoothing technique in signal processing. It is preferred to a rolling mean since it is less sensitive to outliers, which results in better characterization of local trends. Note that there is a truncation for the beginning and end values of the dataset through this method. However, this is of no consequence, given that the extrema are not at the beginning and ends of the dataset. The filtered values are put into a new dataset and do not mutate the original dataset. It is important to note that while the standard deviation is utilized to measure the dispersion, it does not imply that the dynamics is inherently stochastic.

\section{Lifetimes of Four-planet Systems} \label{sec:results}

We integrated more than 40,000 four-planet systems utilizing the sets of initial longitudes detailed in Section \ref{subsec:initial longitudes}. Each set of integrations was started slightly wide of the two-planet orbital stability limit of $\beta = 2\sqrt{3} \approx 3.4641$, equivalent to an initial period ratio between neighboring planets of $1.0677$. The sets of integrations with SL09, Aligned, and Random longitudes are integrated up to $\beta = 8.500$ with a resolution of $0.001$. The three sets of longitudes approaching conjunction and the set with Hexagonal initial longitudes are only integrated to $\beta = 8.200$, just before the steep increase in lifetimes found for the other three sets of longitudes. The SL09 simulations are further integrated up to $\beta = 10$ and $\beta = 8.7$ for Aligned longitudes at a resolution of $0.01$. Additionally, Randomand SL09 longitudes were integrated at a resolution of $\beta = 0.0005$ up to $\beta = 8.2$. All integrations are stopped after $10^{10}$ years if the system survived for this amount of time, but no simulation with $\beta \leq 8.5$ did so. Results are compared with similar integrations for three-planet systems in \cite{lissauer_gavino_2021}, five-planet Random longitudes systems in \cite{Obertas}, and five-planet SL09 and Aligned longitude systems at a resolution of $0.001$ integrated to $\beta = 8.5$. 

\begin{figure}
\vspace{-40mm}
\includegraphics[width=\linewidth]{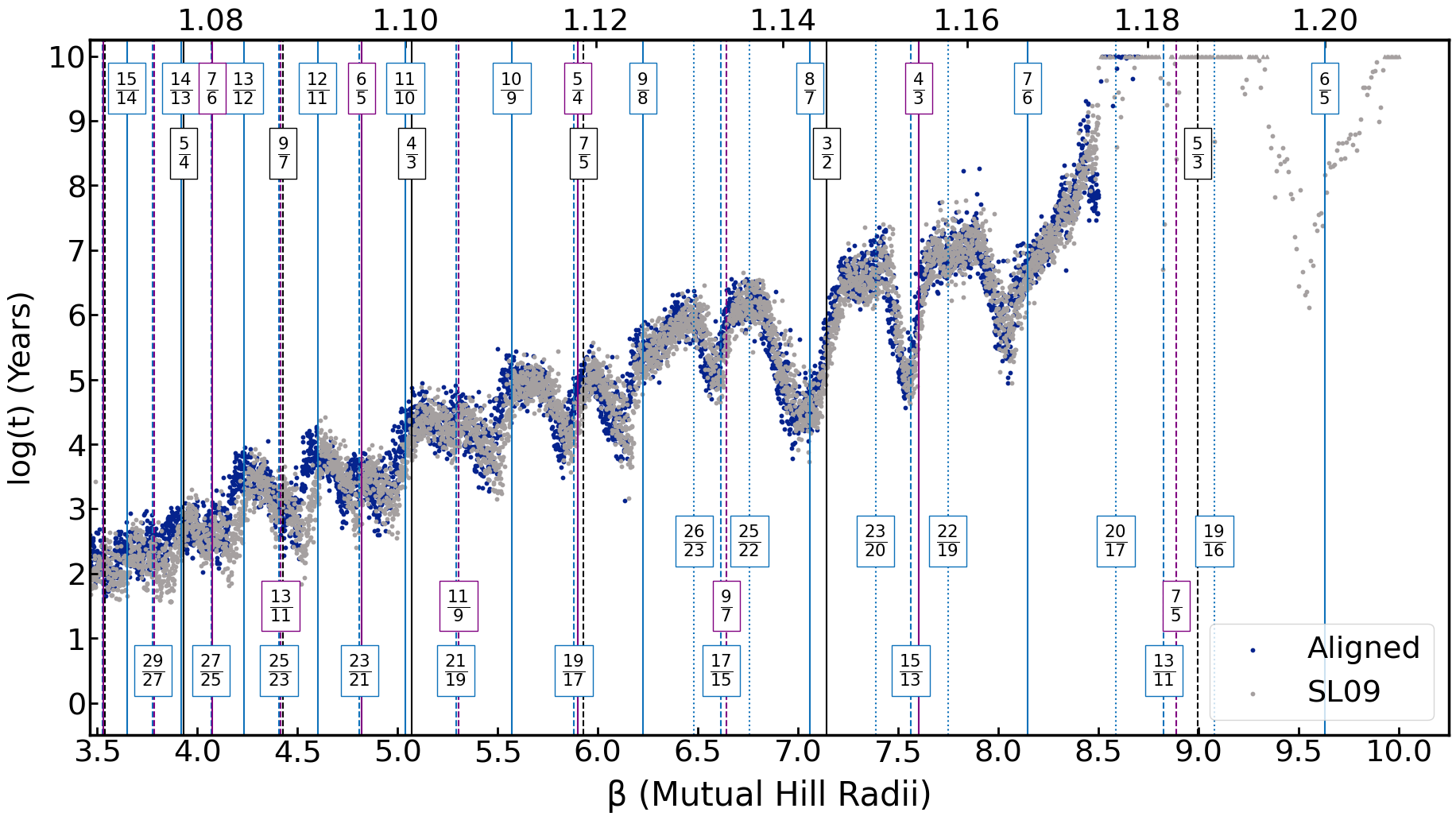}
\includegraphics[width=\linewidth]{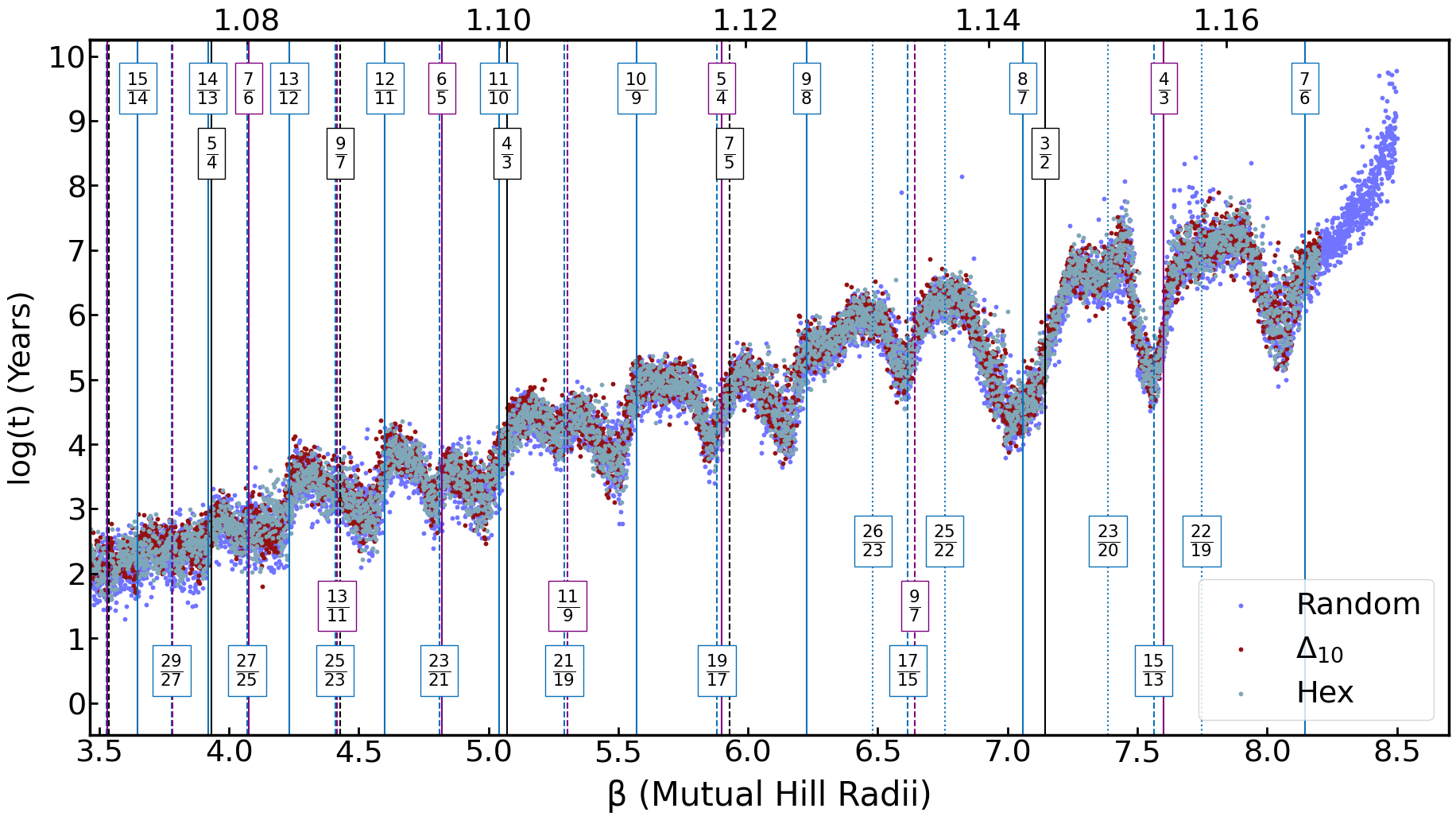}
\caption{The top panel shows the lifetimes of four-planet systems with SL09 (grey) and Aligned (blue) initial longitudes with MMRs overlain.  The bottom panel displays the simulated lifetimes for the Random (purple), $\Delta_{10}$ (red), and Hexagonal (light blue) systems. Here and elsewhere in this work, the order of the overlaid plots corresponds to the order in the legend, as can be seen here with the grey SL09 points on top of the blue Aligned points. Period ratios between neighboring planets are marked above each panel. All results are plotted with a resolution of 0.001 $R_H$ (mutual Hill radii, see Equation \ref{eq:beta_def}) for systems with $\beta \leq 8.5$, above which system lifetimes are plotted at a resolution of 0.01 up to $\beta = 10$ for SL09 longitudes and $\beta = 8.7$ for Aligned longitudes. Likewise, the resolution was 0.001 $R_H$ for Random longitudes up to $\beta = 8.5$, while $\Delta_{10}$ and Hexagonal were completed up to $\beta = 8.2$. Upward pointing triangles represent systems which remained stable at  $10^{10}$ years, when the integrations were terminated. The vertical lines correspond to the regions of first-order (solid lines), second-order (dashed lines), and third-order (dotted lines) MMRs of neighboring planets (blue), pairs with a single planet orbiting between (purple), and the innermost planet with the outermost planet (black). Note that some resonances near the two-planet stability limit do not have callout boxes. No clear pattern of reduction was observed in the lifetimes of systems near second-order resonances between the innermost and outermost planets nor third-order resonances between neighboring planets for systems more tightly-spaced than $\beta \approx 8.52$, so these resonances are not displayed for small values of $\beta$. Third-order resonances between neighbors are labeled beginning at $26/23$, with the first clear effect observed at $20/17$. 
 \label{fig:SL09Align}}

\end{figure}

Figure \ref{fig:SL09Align} shows the simulated lifetimes of four-planet systems as a function of initial orbital separations given in mutual Hill radii.  Systems with Aligned and SL09 initial longitudes are shown in the top panel and highlight the differences in lifetimes for widely disparate initial longitude prescriptions.  The bottom panel shows results for Random, Hexagonal, and $\Delta_{10}$ longitudes, representing three well-separated sets of initial longitudes. Following other studies, we find an exponential trend in system lifetimes with semimajor axes spacing in terms of mutual Hill radii. There are local reductions in system lifetimes that correlate with the locations of MMRs, with the most prominent being first-order resonances, as discovered by \cite{SL09}. We also find that there is a reduction of system lifetimes due to second-order resonances identified by \cite{Obertas}.  We even see evidence for third-order resonances destabilizing system stability, though such MMRs are much more localized in terms of the range of spacings affected than the first- and second-order MMRs; see \ref{sec:third-order Resonances} for details. \\
\subsection{SL09 Longitudes} \label{sec:SL09 longitudes}

Our most extensive set of systems use SL09 longitudes and separations in $\beta$ that are multiples of 0.0005 within the range [3.4645, 8.2], 0.001 over the range (8.2, 8.5], and 0.01 over the range of (8.5, 10]. The reduced resolution in the range $\beta > 8.5$ was used due to the high computational costs of the long-lived systems therein. In total, 9922 simulations were performed, of which 71 were halted at $10^{10}$ years, with the most closely-spaced such system found at $\beta = 8.51$. 

Despite the limited resolution in the range [8.5, 10], we observe the influence of a broad and significant reduction in system lifetimes near the $7/6$ MMR between neighboring planets.  System lifetimes near this local minimum are at least four orders of magnitude less than most similarly-spaced systems.  Additional resonance-induced reductions between neighboring planets are observed for the second-order $13/11$ MMR and third-order $19/16$ and $20/17$ MMRs. The lifetime reduction stemming from a second-order $7/5$ MMR between non-neighboring planets with one planet orbiting between resonating pairs is also evident. We find that both the affected range of planetary spacings and the extent of lifetime reduction diminish for higher order MMRs and MMRs between non-neighboring planets, as discussed further in Section \ref{subsec: Destabilizing Effects of MMRs}.  We do not see evidence for the second-order $5/3$ MMR between inner and outer planets, though this may be due to the combined expected weakness of this resonance and the fact that our integrations were halted at $10^{10}$ years. 

\subsection{Aligned Longitudes} \label{sec:aligned longitudes}

Systems with Aligned longitudes were investigated with separations in $\beta$ that are multiples of 0.001 spanning the range [3.465, 8.5] and 0.01 over the range of (8.5, 8.7]. In total, 5056 simulations were performed, of which 14 were halted at $10^{10}$ years. The most tightly-spaced Aligned system to survive for $10^{10}$ years had $\beta=8.52$, only slightly more widely spaced that the counterpart for SL09 longitudes with $\beta=8.51$.

Notably, most extrema in system lifetimes appear to occur at slightly smaller orbital separation than those found for the SL09 longitudes set of runs (e.g., the local minima near $\beta \approx 4.75$). 
This ``phase shift'' of the ensemble of Aligned systems' lifetimes compared to that of SL09 longitudes is caused by asymmetric energy exchanges between the planets near the start of the integrations. As the planets diverge from their initial alignment, energy is transferred outwards, causing the semimajor axes to shift toward wider spacing.  This is explained and demonstrated in greater detail in Section \ref{subsec: Mechanism Responsible for Phase Shifting}. Examining four-planet systems in Table \ref{tab:trend_line} shows that the Aligned systems' (and the very similar $\Delta_1$ systems) best-fit trend lines have the smallest slopes (i.e., the weakest dependence on initial separation) and the largest separation needed for a $10^{10}$ year lifetime, implying typically lower stability for widely-spaced systems. 

\subsection{Planets At or Near Conjunction} \label{sec:Delta Separation longitudes}
The phase shift between extrema in lifetimes of the two sets of systems discussed in Section \ref{sec:SL09 longitudes} and Section \ref{sec:aligned longitudes} motivated us to integrate systems in a configuration where the initial conjunctions of all planet pairs occur during the first orbit and nearly simultaneously. The goal was to ascertain whether Aligned systems are a unique special case or simply the most extreme member of a trend with decreasing longitudinal spacing.  Neighboring planet longitudes were all initially separated by either $\sim10^\circ$ (following \citealt{lissauer_gavino_2021}), $\sim5^\circ$, or $\sim1^\circ$ (see Table \ref{tab:longitudes_table} for precise values), which are referred to as $\Delta_{10}$, $\Delta_5$, and $\Delta_1$ for simplicity. We integrated these systems over the interval $3.465\leq \beta \leq 8.200$ with a resolution of 0.001. 

The Aligned longitudes, $\Delta_0$, demonstrate the most extreme case of initial conjunction. Each increase in initial longitudinal separation increases the distances between the planets in the system, which allows for greater symmetry in the energy transfers between planets near their first close passage. As can be seen in Table \ref{tab:trend_line}, the extrema of system lifetimes of the $\Delta_{1}$ initial longitudes are close to those of the Aligned longitudes, whereas $\Delta_{5}$ are on average shifted relative to the Random set by only about half as much, and the shift for $\Delta_{10}$ is reduced by another factor of about two, to a value only a few times as large as the uncertainty.

\subsection{Hexagonal Longitudes} \label{sec:Hexagonal longitudes} 
Hexagonal longitudes represent another set of well-spaced initial longitudes.  We expect them to have similar trends in system lifetime as the SL09 longitudes, and we integrate such systems over the interval $3.465\leq \beta \leq 8.200$ for comparison. Hexagonal longitudes have greater stability at smaller separations than SL09 longitudes, as seen in Table \ref{tab:trend_line}, with shifted intercepts $c'$, see Equation 8, of 2.221 and 2.118 for Hexagonal and SL09 longitudes, respectively. However, SL09 longitudes show a marginally greater increase in lifetime with mutual Hill radius. Together, these affects result in the projected spacing for system lifetime of 10 gigayears based upon lifetimes over the range of separations studied for Hexagonal longitudes for these two sets of starting longitudes being approximately equal. This demonstrates that the SL09 longitudes are not overly contrived to produce anomalous lifetimes when compared against another well-spaced longitude set.

\subsection{Random Longitudes} \label{sec:random longitudes}
Systems with each of the planets beginning at randomly-selected longitudes that were drawn independently for each orbital spacing are integrated over the interval $3.465\leq \beta \leq 8.500$ with a resolution of 0.0005. Specifically, the randomly selected longitudes are generated by drawing from a continuous uniform distribution with bounds of $[0.0, 1.0)$ scaled by a factor of $2\pi$. As can be seen in Table \ref{tab:trend_line}, on average, the Random longitude systems have lifetimes that are between the SL09 and Aligned longitudes, but trended significantly closer to the SL09 longitudes. This result is quite intuitive, since, on average, one would generally expect most systems to be somewhat well-separated when initial longitudes are randomly drawn from a uniform distribution.

\subsection{Destabilizing Effects of MMRs} \label{subsec: Destabilizing Effects of MMRs}

Coherent perturbations typically reduce the lifetimes of planetary systems near mean motion resonances, as seen in Figure \ref{fig:SL09Align}. \cite{SL09} demonstrated the destabilizing characteristics of first-order MMRs between neighboring planets on the lifetimes of geometrically-spaced, uniform mass, five-planet systems (see their Figures 1 and 12a). \cite{Obertas} showed that second-order MMRs between adjacent pairs of planets and first-order MMRs between planetary pairs with one planet orbiting between also tend to reduce system lifetimes, though typically over a narrower range in planetary spacing than for first-order MMRs between adjacent planets. \cite{Obertas} found that the dips in system lifetimes associated with first-order MMRs between neighbors are centered at initial period ratios slightly smaller than the nominal resonance location, whereas the dips produced by second-order MMRs are centered near the resonance locations. The cause of this displacement is the tendency for the average period ratios of excited planetary systems with initial period ratios slightly interior to first-order MMRs to increase towards resonant ratios, as explained by \cite{lissauer_gavino_2021}; see also Figure 2 of \cite{EspresateLissauer2001}. \cite{lissauer_gavino_2021} noted that the reductions in system lifetimes by second-order MMRs between neighboring planets are also centered inwards of the nominal resonance locations, but the displacement is much smaller than that of the dips caused by first-order MMRs. 

Numerous resonance-induced dips in system lifetimes are present in Figure \ref{fig:SL09Align}. These dips are associated with all three groups of resonances mentioned in the previous paragraph as well as higher-order resonances and resonances between more distant planets. Specifically, we find that first-order resonances with two planets interjacent, second-order resonances between non-adjacent planets, and third order resonances between neighboring planets are associated with dips in system lifetime. The effects of such resonances are not apparent until greater separations, as they are diminished in magnitude with respect to stronger resonances between proximal planets. These weaker resonances require more time (and thus greater distance from stronger resonances) to excite systems to instability. 

As observed by \cite{Obertas} in five-planet systems, we find that second-order resonances have appreciable repercussions on the system stability of four-planet systems. For closely-spaced systems, the effects of these resonances are small, in marked contrast with first-order resonances. However, effects of the perturbations to the system become more prominent in spacial regions that lack first-order resonances as the system lifetimes increase. For example, the $17/15$ between the first and second planet and the $9/7$ between the first and third planets are a proximate pair of second-order resonances that together lead to a large reduction in system lifetimes for $\beta \approx 6.6$. A much smaller reduction is found for the analogous 21:19 and 11:9 pair around $\beta = 5.3$. In contrast, at larger separation, the 7/5 resonance between the first and third planets is observed to cause a dip in lifetimes on its own (Figure \ref{fig:third_order}), a previously-unidentified result. However, the strength of this particular MMR cannot be fully determined because  systems with slightly different initial orbital separations survive for the entire $10^{10}$~years simulated.  

There is clear evidence of first-order resonances between the inner and outermost planets reducing lifetimes of some SL09 systems. As shown in Figure \ref{fig:SL09Align}, evidence of such resonances isn't apparent for closely-spaced system such as the $5/4$ and $4/3$ MMR. However, there is a unique extremum feature near $\beta \approx 7.0$. Compared to other local extrema, this decline in system lifetimes is significantly wider than other troughs. The $3/2$ resonance between innermost and outermost planets appears to add constructively with the $8/7$ resonance between neighboring planets, producing a  destabilizing effect over a broad range in planetary spacing. As will later be seen in Figure \ref{fig:period shift}, there is a lessened effect on system stability for MMRs between non-neighboring planets when compared to a first-order resonance between neighbors, but the system destabilization of first-order resonances between more distant planets is not inconsequential. We find no direct manifestations of lifetime reductions that correspond with second-order resonances between the innermost and outermost planets. Initial spacings near these resonances are nearly coincident with other resonances or are in a region where lifetimes are greater than $10^{10}$ years precluding a clear determination of the effects of such resonances.   

An unexpected result is the significant reduction in system lifetime found in the immediate vicinity of third-order resonances between neighboring planets at orbital separations where lifetimes of most systems are comparable to our maximum simulation time, 10 gigayears, specifically near the $20/17$ and $19/16$ resonances. To further confirm the importance of these resonances, integrations were performed at high resolution spacing of initial orbital separation in the neighborhood of the $20/17$ and $19/16$ resonances, covering $8.80\leq \beta \leq 8.86$ and $9.07\leq \beta \leq 9.09$, respectively. The results are displayed in Figure \ref{fig:third_order} within \ref{sec:third-order Resonances}. 
Other third-order resonances at somewhat smaller separations are also marked in Figure \ref{fig:SL09Align}, however, the effect of such resonances on shorter-lived systems is practically imperceptible. Only for the longer-lived systems do the persistent, repeated effects of these weaker MMRs have an observable effect on the stability of the overall system.


\begin{table}[H]
    \vspace{-20mm}
    \caption{Log-linear fits of system lifetime as a function of initial orbital separation for different planetary multiplicity, initial longitudes and range in separation. Note that 71 four-planetary integrations with SL09 longitues, all with  $\beta >8.5$, were terminated when the system reached $10^{10}$ years; a lifetime of $10^{10}$ years was assumed for these systems in the fit for SL09 longitudes out to $\beta = 10$. Three-planet data from \cite{lissauer_gavino_2021} and five-planetary Random longitude data from \cite{Obertas} are for systems integrated for $10^{10}$ years or until an orbit crossing occurred. All of the other systems used for the fits presented in this table were integrated up to the point when the system became unstable.
    $^\dagger$ All three-planet data herein were retrieved from \cite{lissauer_gavino_2021}. 
    $^\ddagger$ All five-planet Random longitude data herein were retrieved from \cite{Obertas}; note that the ranges and resolutions are approximate since the separations were drawn from a uniform distribution on the interval specified. }
    \begin{center}
    \tiny
    \begin{adjustbox}{angle=90}
    \begin{tabular}{ccccccccc}
    \# of Planets & Longitudes & Resolution & Range & $b'$ & $c'$ & $\beta$ at $\log t_c = 10$ & $\sigma_{\rm exp}$ & $\sigma_{\rm local}$ \\
    \midrule
    \hline

    3     & Random$^\dagger$ & 0.0005 & [3.4645, 5.979] & 1.422 $\pm$ 0.010 & 2.078 $\pm$ 0.014 	& 8.01 $\pm$ 0.07 & 0.497 & 0.376\\
    3     & SL09$^\dagger$ & 0.001 & [3.465, 6.300] & 1.383 $\pm$ 0.012 & 2.063 $\pm$ 0.020 & 8.24 $\pm$ 0.09 & 0.540 & 0.312  \\
    3     & SL09$^\dagger$  & 0.01  & [3.47, 7.15] & 1.461 $\pm$ 0.037 & 1.998 $\pm$ 0.079 & 7.85 $\pm$ 0.26 & 0.754 & --- \\

    \hline

    4 	& Random & 0.0005 & [3.4645, 5.979] & 1.146 $\pm$ 0.007 & 2.029 $\pm$ 0.011  & 9.98 $\pm$ 0.08 & 0.386 & 0.247 \\
    4     & SL09  & 0.001 & [3.465, 6.300] & 1.137 $\pm$ 0.009 & 1.967 $\pm$ 0.015 & 10.11 $\pm$ 0.10 & 0.412 & 0.226 \\
    4 & SL09 & 0.01 & [3.47, 7.15] & 1.064 $\pm$ 0.025 & 2.066 $\pm$ 0.054 & 10.71 $\pm$ 0.31 & 0.513 & --- \\

    4     & Aligned & 0.001 & [3.465, 8.200] & 0.980 $\pm$ 0.006 & 2.328 $\pm$ 0.016 & 11.37 $\pm$ 0.08 & 0.538 & 0.226 \\
    4     & $\Delta_1$ & 0.001 & [3.465, 8.200] & 0.993 $\pm$ 0.006 & 2.351 $\pm$ 0.016 & 11.19 $\pm$ 0.08 & 0.542 & 0.231 \\
    4     & $\Delta_5$ & 0.001 & [3.465, 8.200] & 1.001 $\pm$ 0.006 & 2.311 $\pm$ 0.016 & 11.14 $\pm$ 0.08 & 0.539 & 0.222\\
    4     & $\Delta_{10}$ & 0.001 & [3.465, 8.200] & 1.005 $\pm$ 0.006 & 2.293 $\pm$ 0.016 & 11.12 $\pm$ 0.08 & 0.543 & 0.221 \\
    4     & Hex   & 0.001 & [3.465, 8.200] & 1.021 $\pm$ 0.006 & 2.221 $\pm$ 0.016 & 11.01 $\pm$ 0.08 & 0.540 & 0.230 \\
    4     & SL09  & 0.001  & [3.465, 8.200] & 1.029 $\pm$ 0.006 & 2.118 $\pm$ 0.016 & 11.02 $\pm$ 0.08 & 0.547 & 0.229\\
    4 & SL09 & 0.001 & [3.4645, 8.1995] & 1.033 $\pm$  0.006 & 2.109 $\pm$  0.016 &	11.00 $\pm$  0.08 & 0.551 & 0.233 \\

    4 	& SL09 	& 0.0005 & [3.4645, 8.2000] & 1.031 $\pm$ 0.004 & 2.114 $\pm$ 0.011 &	11.01 $\pm$ 0.06 & 0.549 & 0.234\\

    4 & Random & 0.0005 & [3.4645, 8.2000] & 1.012 $\pm$ 0.004 & 2.198 $\pm$ 0.011 & 11.13 $\pm$ 0.06 & 0.547 & 0.257\\

    4     & Aligned & 0.001 & [3.465, 8.500] & 1.019 $ \pm$ 0.005 & 2.263 $\pm$ 0.015 & 10.99 $\pm$ 0.07 & 0.548 & 0.226\\  
    4     & SL09  & 0.001  & [3.465, 8.500] & 1.069 $\pm$ 0.005 & 2.052 $\pm$ 0.016 & 10.67 $\pm$ 0.07 & 0.561 & 0.228\\
    4     & Random & 0.001 & [3.465, 8.500] & 1.049 $\pm$ 0.005 & 2.139 $\pm$ 0.016 & 10.79 $\pm$ 0.07 & 0.563 & 0.257 \\

    4     & SL09  & 0.01 & [3.47, 10.00] & 1.207 $\pm$ 0.018 & 1.837 $\pm$ 0.066 & 9.63 $\pm$ 0.20 & 0.847 & ---\\

    \hline

    5     & Random$^\ddagger$ & 0.0005 &(3.465, 5.979) & 1.082 $\pm$ 0.007 & 1.939 $\pm$ 0.011 & 10.65 $\pm$ 0.08 & 0.378 & 0.214\\
    5     & SL09 & 0.001 & [3.465, 6.300] & 1.079 $\pm$ 0.009 & 1.870 $\pm$ 0.015 & 10.74 $\pm$ 0.11 & 0.407 & 0.205\\
    
    5     & SL09 & 0.01 & [3.47, 7.15] & 0.983 $\pm$ 0.025 & 2.005 $\pm$ 0.054 & 11.66 $\pm$ 0.37 & 0.520 & --- \\

    5     & Aligned & 0.001 & [3.465, 8.500] & 0.960 $\pm$ 0.005 & 2.161 $\pm$ 0.015 & 11.78 $\pm$ 0.08 & 0.539 & 0.198 \\
    5     & SL09  & 0.001 & [3.465, 8.500] & 0.995 $\pm$ 0.005 & 1.976 $\pm$ 0.015 & 11.54 $\pm$ 0.08 & 0.545 & 0.204 \\
    5     & Random$^\ddagger$ & 0.0005 & (3.465, 8.500) & 0.973 $\pm$ 0.004 & 2.060 $\pm$ 0.011 & 11.72 $\pm$ 0.06 & 0.539 & 0.207 \\

    5 & Random$^\ddagger$ & 0.0005 & (3.465, 10.000)	& 1.087 $\pm$ 0.004 &  1.878 $\pm$ 0.014 & 	10.66 $\pm$ 0.05 & 0.783 & 0.239 \\
        
    \end{tabular}%
    \end{adjustbox}
\end{center}

    \label{tab:trend_line}%
\end{table}%

\section{Comparison with Lifetimes of Three-planet and Five-planet Systems} \label{sec:345}

In this section, we place the ensemble of lifetimes of four-planet systems presented in Section \ref{sec:results} into the broader context of closely-spaced multi-planet systems by comparing them to the lifetimes of analogous systems with three or five planets. For this comparison, we use previous data sets where available, filling in the gaps with new integrations as needed. For three-planet systems, we consider SL09 and Random longitudes obtained from \cite{lissauer_gavino_2021}, as their results for Aligned longitudes show a major spike in lifetimes near $\beta = 5.17$ that complicates quantitative comparisons with four-planet systems starting at Aligned longitudes. For five-planet systems, we use the extensive set of Random longitudes lifetimes obtained from \cite{Obertas}, and perform more limited sets of new integrations using SL09 and Aligned longitudes to compare with similar longitudes in four-planet systems.  

As seen in Figure \ref{fig:3_4_5} and quantified in Table \ref{tab:trend_line},  the lifetimes of four-planet systems are generally intermediate between those of three- and five-planet systems, but closer to five-planet systems based upon the trendlines seen in Table \ref{tab:trend_line}. Specifically, when adjusting the linear regression of the logarithm of system lifetime data following \cite{QuarlesLissauer2018}, the log-lifetime slopes with respect to the initial spacing for systems with SL09 starting longitudes over the range $\beta \in$ [3.465, 6.300] are $1.383$, $1.137$, and $1.079 $ for three, four and, five planets, respectively. There is a diminishing and decreasing trend in the log-lifetime slopes with respect to the number of planets in the system, as also seen in \cite{SL09}. However, the log-lifetime slopes of the four-planet systems are closer to the five-planet systems than three-planet systems.

\begin{figure}
\centering
    \centering
    \includegraphics[width=\linewidth]{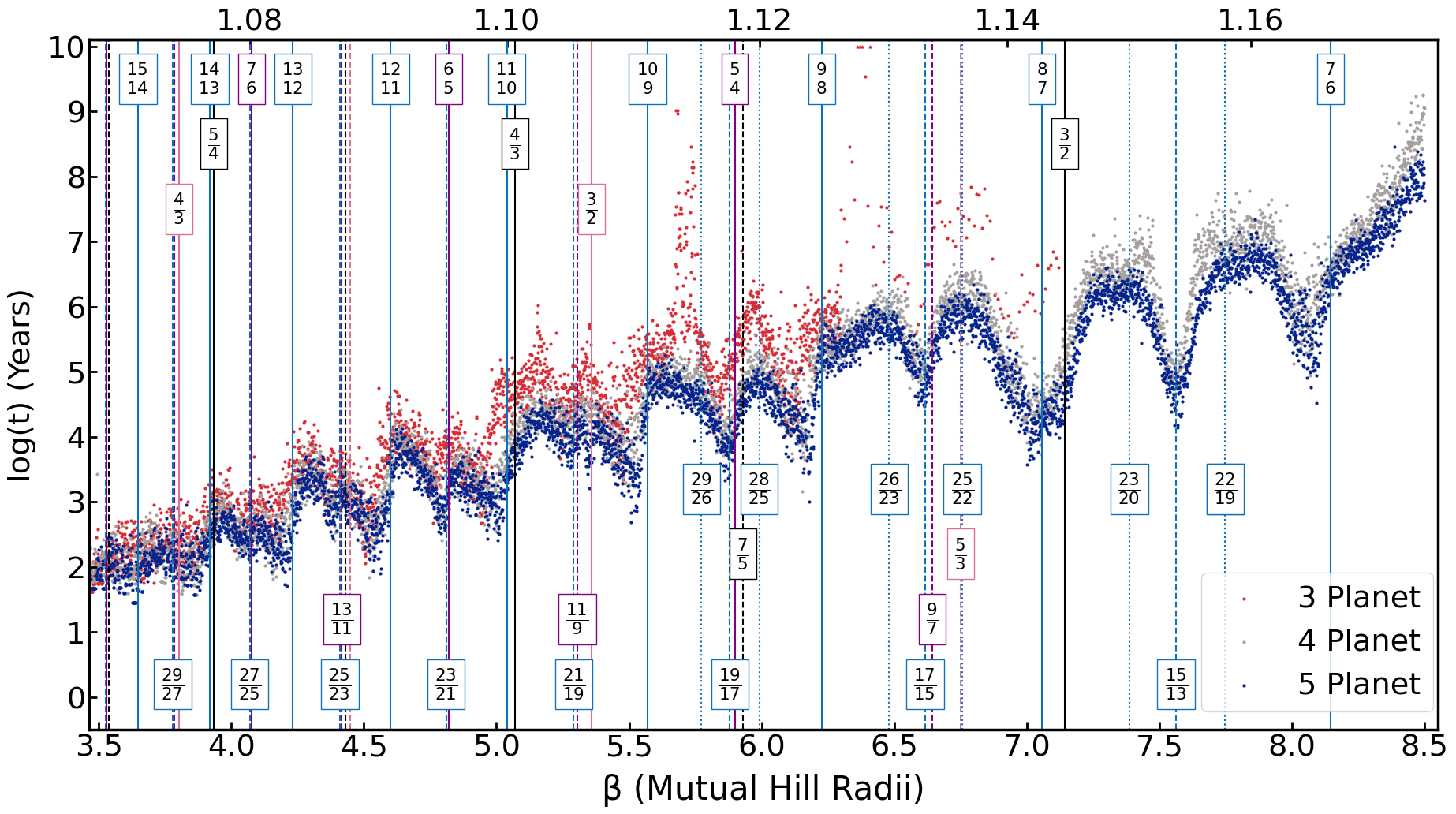} \label{fig:3_4_5} \\
    \includegraphics[width=\linewidth]{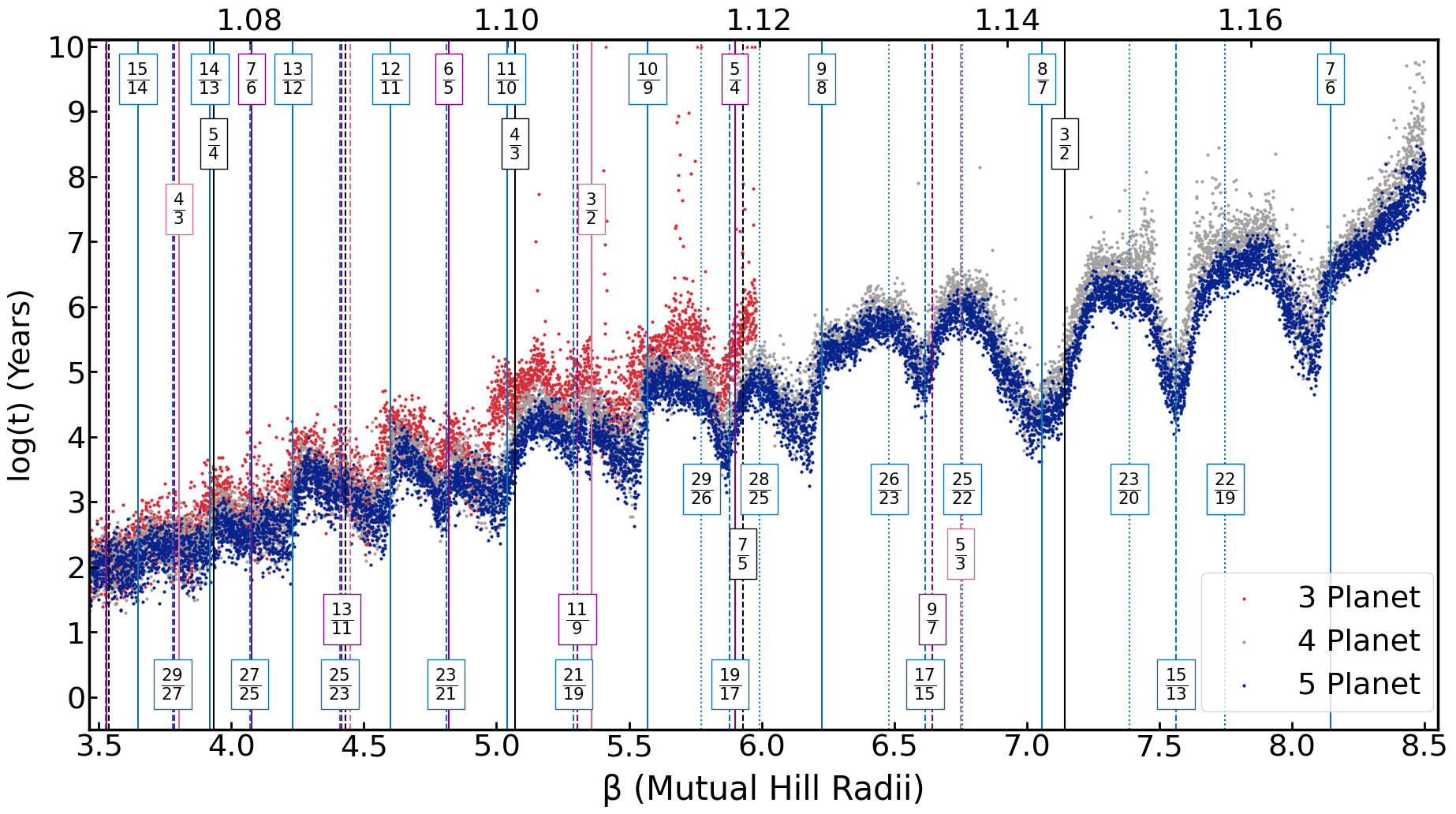} \label{fig:3_4_5_rand} \\
    \caption{Overlain plots of the lifetimes of three- (red dots), four- (gray dots), and five-planet (blue dots) systems with SL09 (Top) and Random (bottom) initial longitudes at a resolution of 0.001 in mutual Hill radii. Red points are plotted first, followed by grey points, and finally blue points.  The three-planet lifetimes are from \cite{lissauer_gavino_2021}. Additionally, five-planet Random lifetimes were retrieved from \cite{Obertas}. Initial orbital period ratios of neighboring planets are listed on the top axes. The vertical lines correspond to the regions of first-order (solid lines), second-order (dashed lines) and third-order (dotted lines) MMRs of neighboring planets (blue), pairs with one intermediate planet orbiting between (purple), pairs with two intermediate planets orbiting between (black), and pairs with three intermediate planets orbiting between (pink) for the five-planet systems.}
    \label{fig:3_4_5}
\end{figure}

\begin{figure}
    \centering
    \includegraphics[width=\linewidth]{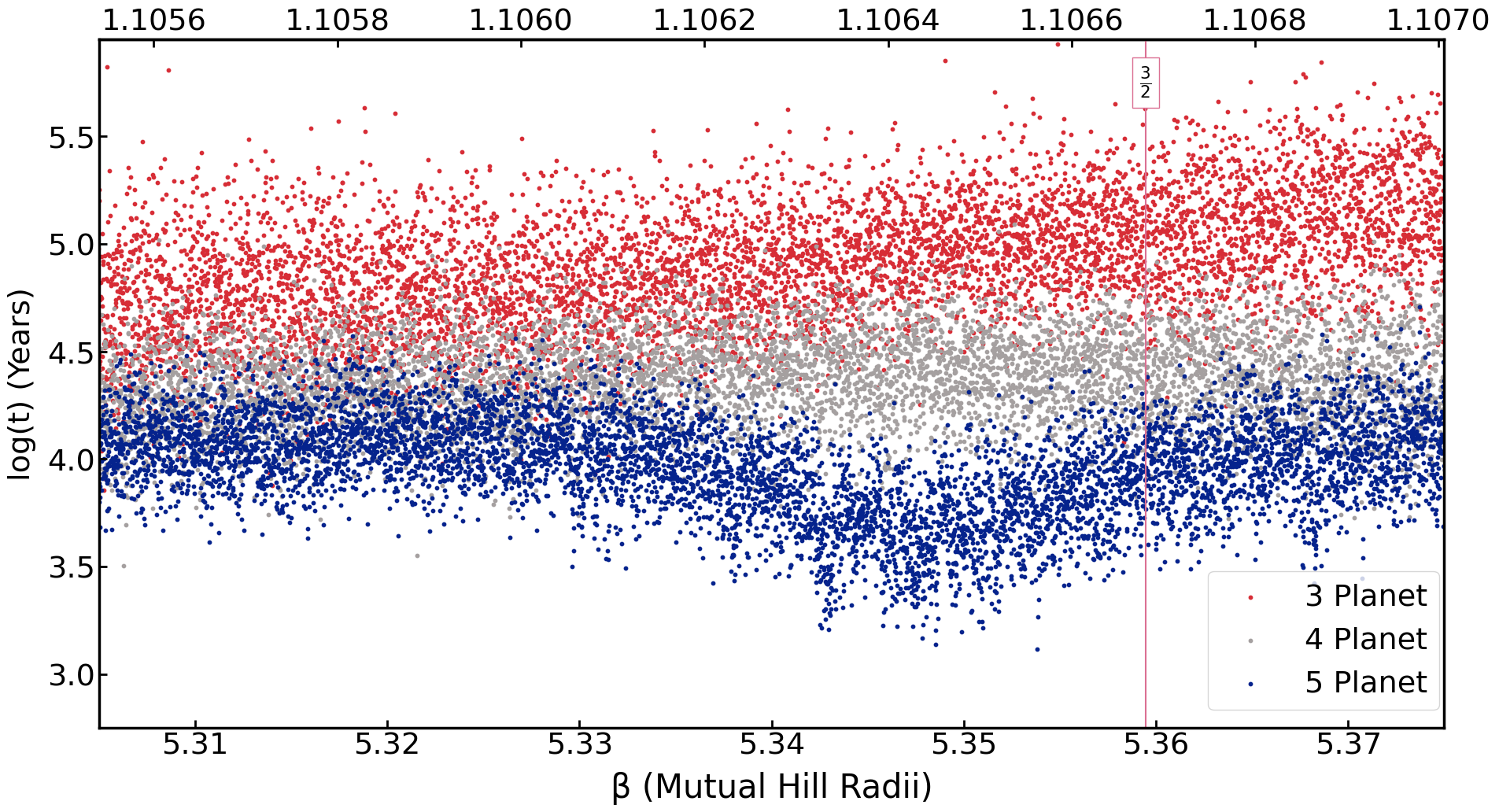}
    \caption{Lifetimes of three-, four-, and five-planet systems with SL09 initial longitudes and orbital separations at a resolution of $10^{-6}$ mutual Hill radii are superimposed in the region corresponding to a 3/2 first-order resonance between the first and fifth planets in five-planet systems. 
    Within this small window, typical  lifetimes of the three- and four-planet systems increase roughly linearly with orbital separation. However, there is a significant decrease in system lifetimes in the five-planet systems close to and somewhat narrow of the 3/2 first-order resonance between the first and fifth planets, a resonance not present for systems with fewer planets. 
    }
    \label{fig:5_planet_zoom}
\end{figure}

There are some characteristics that pertain to systems with a given number of planets that do not appear for other systems. A key difference evident in Figure \ref{fig:3_4_5} is that systems with four or more planets lack the ``spikes" in system lifetimes observed in three-planet systems by \cite{lissauer_gavino_2021}. Another trend is that additional resonances are observed to reduce the lifetimes of systems with more planets. For example, comparing Figures \ref{fig:SL09Align} and \ref{fig:3_4_5}, there is a small, but significant, narrow dip in five-planet system lifetimes at $\beta \approx 5.35$ that corresponds with the $3/2$ MMR between the first and fifth planets in the system. This dip is not apparent in the three- and four-planet systems that lack this resonance. Figure \ref{fig:5_planet_zoom} provides a closer look at this region. 

Additional differences between planetary systems are found when examining the amount of scatter in systems of different planetary multiplicity and initial longitudes, quantified with the dispersion metrics $\sigma_{\rm exp}$ and $\sigma_{\rm local}$ discussed in Section \ref{sec: Dispersion Metrics}. We apply these metrics to various sets of simulated systems and ranges of orbital separation and present results in Table \ref{tab:trend_line}. The dispersion metrics of the three-planet systems are substantially greater than those calculated for the four-planet and five-planet systems with the same initial longitudes and over the same ranges. Three-planet systems show ``spikes'' in  lifetimes for a significant fraction of spacings at planetary spacings greater than $\beta \approx 6.3$, with several systems terminated at $10^{10}$ years, as shown in Figure \ref{fig:3_4_5}.  Thus, four- and five-planet systems should only be compared with three-planet systems over a range in $\beta$ of $\left[3.465, 6.300 \right]$. This holds true over small/local and large ranges of planetary spacings as well as for both SL09 and Random longitudes. Similarly, four-planet dispersion metrics for SL09 and Random longitudes are greater than both metrics for the five-planet systems. However, the differences between three- and four-planet systems are greater than those between four- and five-planet systems.  For example, systems with SL09 longitudes with spacings $\ 3.465 \leq \beta \leq 6.300$ have $\sigma_{\rm exp}$ equal to 0.540, 0.412, and 0.407 for three-, four-, and five-planet systems, respectively.  The same systems have $\sigma_{\rm local}$ equal to 0.375, 0.247, and 0.196 for three-, four-, and five-planet systems, respectively. With a comparison between the four- and five-planet systems over the ranges of $\left[3.47, 7.15 \right]$ and $\left[3.465, 8.500 \right]$, the dispersion metrics for four-planet and five-planet are relatively close. Overall, we find that four-planet systems behave more similarly to five-planet systems than three-planet systems in terms of regression fitting parameters, lack of ``spikes'' in system lifetime, and dispersion metrics.

Despite the differences observed in dispersion, long-lived spiking behavior in three-planet systems, and regression line fitting parameters, all sets of systems show dips in lifetimes that correlate with mean motion resonances. One caveat is that three-planet systems lack some of resonances mentioned in Section \ref{subsec: Destabilizing Effects of MMRs} simply because parings in those resonances do not exist in the system.  Under most circumstances, resonances between planets in three-, four-, and five-planet systems tend to perturb the systems toward instability. Perturbations towards stability are present, but are restricted to a small fraction of parameter space as determined by \cite{GavinoLissauer25}. Dynamically, the effects of such perturbations can be seen in changes in the semimajor axis and eccentricity of the planetary orbits. Larger initial separation between semimajor axes of planets generally correlate with increased stability whereas larger initial eccentricity, as studied by \cite{GRATIA2021114038}, tends to reduce the stability of the system.

\begin{figure}
    \centering
    \includegraphics[width=\linewidth]{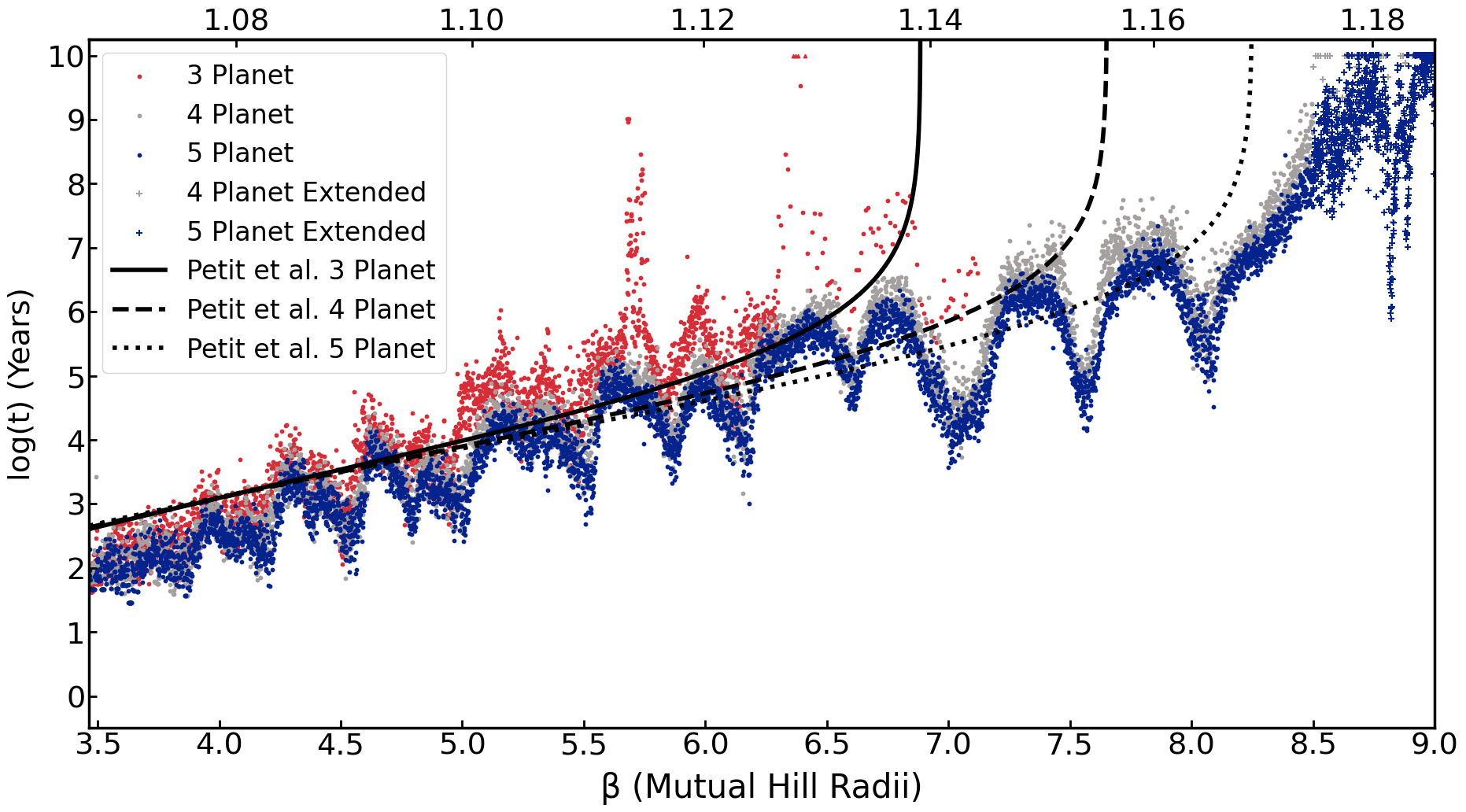}
    \caption{Lifetimes of three-, four-, and five-planet systems with SL09 longitudes from \cite{lissauer_gavino_2021} and this work, respectively are compared against the theoretical system lifetimes of \cite{petit2020path}. Four- and five-planet systems are plotted at differing resolutions past a mutual Hill radius of 8.5, denoted by a $+$, to demonstrate when 10 gigayears is reached. For the five-planet extension, utilizing Randomlongitudes, the results from \cite{Obertas} are used.
    }
    \label{fig: petit}
\end{figure}

\textcolor{black}{We compare our numerical results against the theoretical results of \cite{petit2020path} in Figure \ref{fig: petit}. For the three-planet case, it was shown that, when considering MMRs, the problem can be reduced to unidimensional Chirikov diffusion. However, this is not extendable to systems with greater number of planets except by using a scaling factor to approximate the effects of the other planets. We use a factor of $K=1.5$, midway between $K=1$ for three-planet systems and $K=2$ for five-planet systems in \cite{petit2020path}, to represent the four-planet system. We can further numerically confirm that the theoretical results from \cite{petit2020path} can model the system lifetimes of four- and five-planet systems. Notably, however, there is a stark inconsistency near a period ratio of 1.14 for the three-planet case that continues for the four- and five-planet cases. This is due to a failure of the MMR overlap criterion as mentioned in \cite{petit2020path}. Nonetheless, we see further confirmation that the lifetimes of four-planet systems are closer to those of five-planet systems than three-planet systems, as previously discussed.}

\section{Phase-Shifting for Differing Initial Longitudes} \label{phase_shifts}
An unexpected result of our investigations into four-planet systems is the appearance of a ``phase shift'' displacement of lifetime extrema towards less widely-spaced initial separations for systems with initially Aligned longitudes relative to those with Random (as well as most other) initial longitudes. The lifetime displacement between longitude sets is evident by comparing the locations of the resonance-induced dips as seen in Figure \ref{fig:SL09Align}. To investigate the apparent phase-shifting, we use an extrema discovery method detailed in  \ref{subsec:Data Processing}. This method is applied to the relationship between log lifetimes and initial orbital separation for four-planet systems in Section \ref{the aligned sl09 phase shift}, allowing us to determine the locations of the various local extrema labeled in Figure \ref{fig:rollingmedians}. It is visually evident that systems with Aligned initial longitudes have lifetime extrema shifted inward (to smaller $\beta$) compared with lifetime extrema locations for well-separated systems with SL09 longitudes. Additionally, this shifting appears to decrease in magnitude as the planetary spacing ($\beta$) increases.  Upon comparison with other initial longitude prescriptions, a general progressive trend in phase-shifting is observed as initial longitudes approach conjunction. The physical mechanism responsible for the shift is explained in Section \ref{subsec: Mechanism Responsible for Phase Shifting}.

\subsection{Phase Shifts Dependence on Initial Longitudes} \label{the aligned sl09 phase shift}
\begin{figure}
    \centering
    \includegraphics[width=0.9\linewidth]{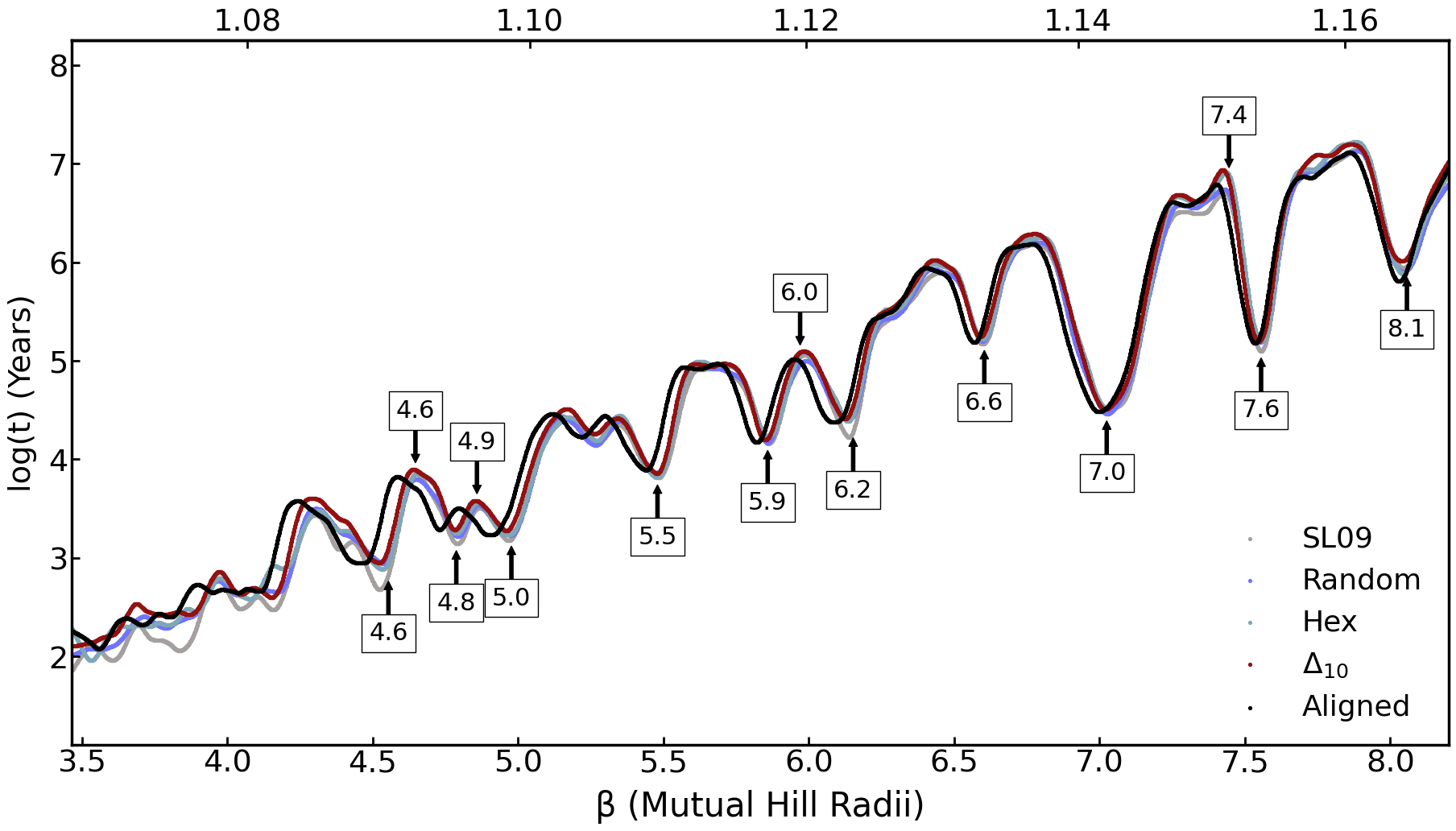}
    \includegraphics[width=0.9\linewidth]{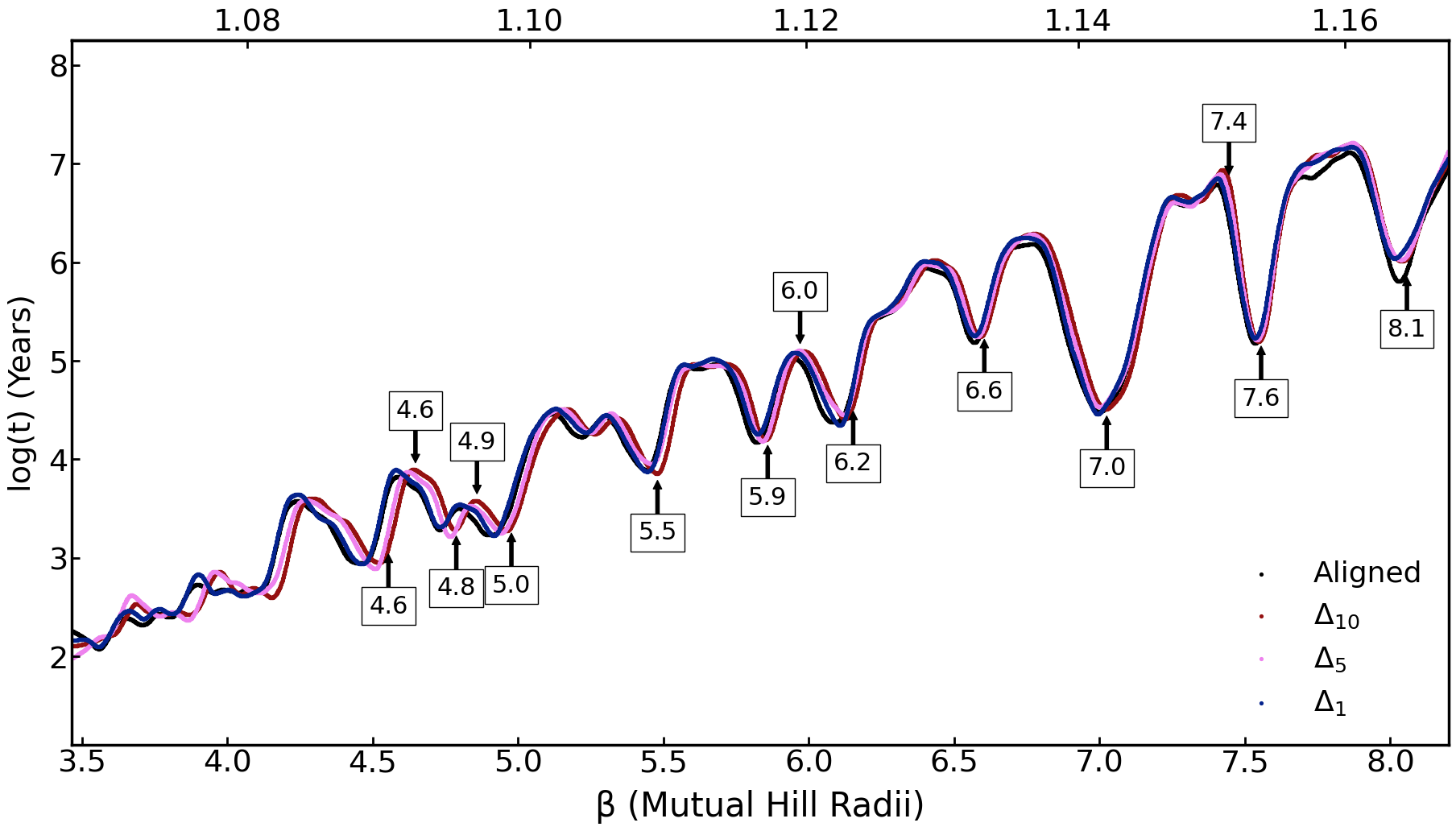}
    \caption{Rolling medians of lifetimes of four-planet systems are plotted against the initial orbital separation of neighboring planets, as detailed in Section \ref{subsec:Data Processing}. The prominent extrema corresponding to Table \ref{tab:4_planet_phase_shift} are marked. For all sets of systems, the smoothed system lifetimes have a resolution of $\beta = 0.001$. As with other figures, the order of the overlaid plots corresponds to the order in the legend, as can be seen with the gray SL09 curve on bottom of the blue Aligned curve in the top panel. In the top panel, the phase-shifting behavior between the four-planetary systems with SL09, Aligned, Random, and Hexagonal longitudes is evident.}
    \label{fig:rollingmedians}
\end{figure}

To analyze the phase-shifting pattern observed, the prominent extrema of the lifetimes vs.~orbital separation for each set of initial longitudes were calculated and tabulated. We restricted our attention to those entries that are common to all sets of longitudes considered, which are labeled in Figure \ref{fig:rollingmedians}. The difference between the locations of the extremum values were used to represent the local phase shift between system lifetime datasets. 

As can be seen in Table \ref{tab:4_planet_phase_shift},  there is no systematic difference between the extrema locations for the SL09, Hexagonal and Random longitudes. In contrast, there is a marked change (relative to the Random set) for the Aligned, $\Delta_{1}$, and $\Delta_{5}$ longitudes. All of the extrema for these starting longitudes have positive phase shifts, and apart from the extrema near $\beta =6$ of the $\Delta_{5}$ longitudes, all shifts exceed the estimated uncertainty. The average phase shifts substantially exceed the estimated uncertainty of $\Delta\beta\approx 0.0042$ (\ref{subsec:Data Processing}).

\begin{table}[]
    \centering
    \caption{Quantitative estimates of phase shifts identified in extrema location for each of the sets with unique starting longitudes relative to the set with Random longitudes, as described in \ref{sec: Phase Shift Quantification}, measured in terms of the difference in planetary spacing at which prominent extrema occur in different systems.  Prominent peaks and troughs are denoted in the first column as ``P'' and ``T'' followed by the approximate planetary spacing location ($\beta$) of the features.  All phase shifts are measured relative to the extrema locations for four-planet Random longitudes at resolution of 0.0005 in $\beta$.  Positive phase shifts represent features that occur in a given dataset at a larger planetary spacing, and negative phase shifts for smaller planetary spacing. The mutual Hill radii values correspond to those in Figure \ref{fig:rollingmedians}. The bottom row of the table displays the mean phase shift estimate for the entirety of the dataset considered, except for the ``Abs. Diff'' column where it represents the mean of the amplitudes of the individual phase shifts. The estimated uncertainty of the method utilized, as detailed in Section \ref{subsec:Data Processing}, is  $\beta\sim0.0042$. The similarity of values in the first pair of rows and that between the third and fourth rows is due to the close proximity to first-order resonance for both the peak and the trough. Note that ``SL09 HR" denotes the set of SL09 system lifetimes at a resolution of 0.0005.  All other columns are comparisons for sets at 0.001 resolution and are, therefore, only recorded to that degree of accuracy.}
    \label{tab:4_planet_phase_shift}
    \footnotesize
    \begin{tabular}{cccc|c|ccccc}
        Extrema & SL09 HR & SL09 Shift & SL09 & Abs. Diff. & Hex & $\Delta_{10}$ & $\Delta_{5}$ & $\Delta_{1}$ & Aligned \\
        \midrule
        \hline
        T-4.6 & 0.0290 & 0.029 & 0.031 & 0.002 & 0.020 & 0.031 & 0.047 & 0.093 & 0.066 \\
        P-4.6 & 0.0290 & 0.029 & 0.031 & 0.002 & 0.020 & 0.031 & 0.047 & 0.093 & 0.065 \\
        T-4.8 & -0.0025 & -0.001 & -0.003 & 0.002 & 0.001 & 0.006 & 0.023 & 0.058 & 0.059  \\
        P-4.9 & -0.0025 & -0.001 & -0.003 & 0.002 & 0.001 & 0.006 & 0.023 & 0.058 & 0.059 \\
        T-5.0 & 0.0130 & 0.026 & 0.010 & 0.016 & 0.011 & 0.017 & 0.036 & 0.062 & 0.078 \\
        T-5.5 & -0.0055 & -0.007 & -0.004 & 0.003 & 0.000 & 0.001 & 0.019 & 0.034 & 0.037 \\
        T-5.9 & -0.0030 & -0.003 & -0.003 & 0.000 & 0.002 & 0.009 & 0.020 & 0.036 & 0.039 \\
        P-6.0 & -0.0140 & -0.015 & -0.014 & 0.001 & -0.016 & -0.011 & 0.004 & 0.020 & 0.019 \\
        T-6.2 & 0.0120 & 0.013 & 0.012 & 0.001 & 0.017 & 0.027 & 0.031 & 0.045 & 0.067 \\
        T-6.6 & 0.0050 & 0.006 & 0.005 & 0.001 & 0.007 & 0.015 & 0.024 & 0.035 & 0.035 \\
        T-7.0 & 0.0020 & 0.004 & 0.001 & 0.003 & 0.006 & 0.008 & 0.027 & 0.033 & 0.030 \\
        P-7.4 & 0.0130 & 0.013 & 0.015 & 0.002 & 0.009 & 0.021 & 0.030 & 0.038 & 0.041 \\
        T-7.6 & 0.0020 & 0.004 & 0.001 & 0.003 & 0.001 & 0.009 & 0.016 & 0.023 & 0.023  \\
        T-8.1 & 0.0080 & 0.004 & 0.012 & 0.008 & 0.016 & 0.016 & 0.021 & 0.045 & 0.027 \\
        \hline
        AVG & 0.0061 & 0.007 & 0.006 & 0.004 & 0.006 & 0.013 & 0.026 & 0.048 & 0.046 \\
    \end{tabular}
\end{table}

The relative positions of extrema for the longitudes approaching conjunction can be seen in the bottom panel of Figure \ref{fig:rollingmedians}. The typical phase shift for the $\Delta_{10}$ longitudes is much smaller than that for the $\Delta_{1}$ longitudes, with the $\Delta_{5}$ longitudes producing intermediate shifts  (Table \ref{tab:4_planet_phase_shift}). The magnitude of the phase shifts are similar for the Aligned longitudes and $\Delta_{1}$. 

As the initial orbits of the planets become farther apart, the phase shift diminishes; i.e., the phase shift is more pronounced for smaller orbital separations (Table \ref{tab:4_planet_phase_shift}).  Notably, there is an exception to this for the extrema near $\beta =6.1$,  which appears to be caused by an anomaly in the Random longitudes that is not present in the other longitude sets.

\subsection{Mechanism Responsible for Phase-Shifting} \label{subsec: Mechanism Responsible for Phase Shifting}
To investigate the cause of the phase-shifting between systems with well-spaced and closely-spaced longitudes, we calculate the mean planetary period and compare the mean period ratios to initial orbital period ratio estimated using Kepler's Third Law, following the methodology presented in Section 5 of \cite{lissauer_gavino_2021}.  In order to observe the aggregate dynamics of the systems, each system is simulated for $10^{3}$ years. During these integrations, a running count of completed and partially completed orbits are counted, along with the simulated time to determine the average period ratios.   

\begin{figure}[H]
\includegraphics[width=0.9\linewidth]{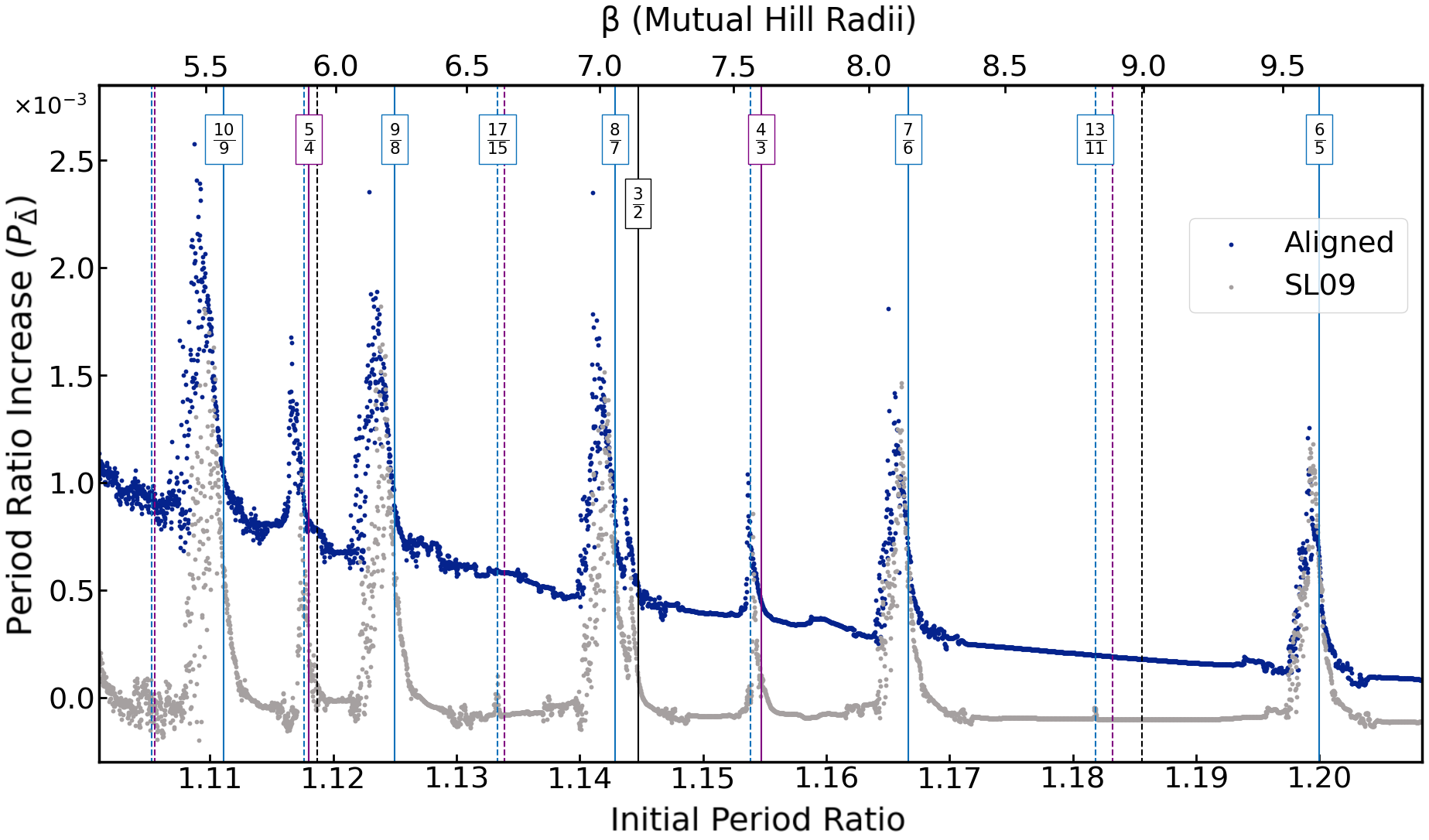}
\vspace{2mm}
\includegraphics[width=0.9\linewidth]{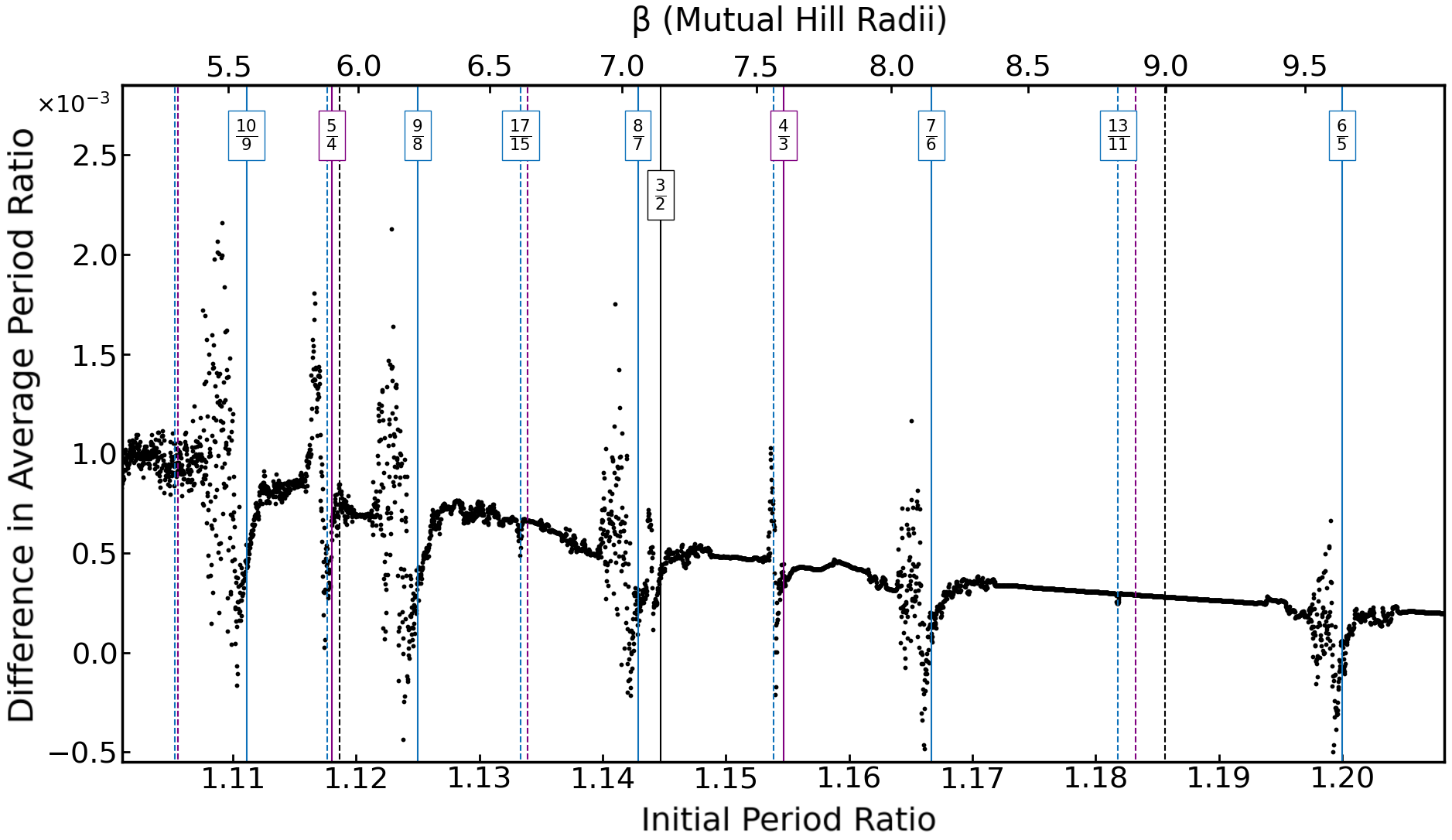}
\caption{The top panel plots the difference between average and initial period ratios of neighboring planets, described in Equation \ref{eq: avg diff period ratio}, over a time span  of $10^3$ simulated years (1000 orbits of the innermost planet) for the Aligned (blue) and SL09 (grey) longitudes. Unlike other figures in this work, the initial period ratio is the bottom axis while the top axis is the separation in mutual Hill radii $(\beta)$. The bottom panel is the difference between the period shifts for Aligned and SL09 initial longitude systems displayed in the top panel. For both panels, the vertical lines correspond to the regions of first-order (solid lines) and second-order (dashed lines) MMRs of neighboring planets (blue), the innermost planet with the third planet (purple), and the innermost planet with the fourth planet (black). The plot shows that mean period ratios between planets are increased more relative to initial values for Aligned systems than for SL09 systems, with the magnitude of the typical difference decreasing as $\beta$ increases.}

\label{fig:period shift}
\end{figure}

The early asymmetric perturbations in Aligned systems result in an increase of the ratios between planetary periods (Fig.~\ref{fig:period shift}). Such an increase perturbs the systems to be analogous to systems with greater initial separation. For illustration, as two planets on circular orbits approach one another, the angular component of the force between them results in an energy transfer from the outer planet to the inner planet. After conjunction, there is then an energy transfer from the inner to the outer planet that, to  first-order, cancels the effects of the previous energy transfer. However, for Aligned initial longitudes, there is an asymmetric energy transport between the planets. Very near conjunction, little work is done because the force between the planets is nearly in the radial direction, perpendicular to their velocities. Because of this, the Aligned and $\Delta_{1}$ longitudes are very similar in characteristics, whereas the phase shift is reduced substantially for the $\Delta_5$  longitudes and further reduced for the $\Delta_{10}$.  

In order to better observe the period shifting and the effect of resonances, the averaged difference in period ratios, $P_{\bar{\Delta}}$, is defined as follows: 

\begin{equation}
\label{eq: avg diff period ratio}
    P_{\bar{\Delta}} = \sqrt[3]{\frac{P_{4,1000}}{P_{1,1000}}} - \sqrt[3]{\frac{P_{4,0}}{P_{1,0}}},
\end{equation}

\noindent where the first subscript denotes the planet. The second subscript of 0 denotes the initial condition, whereas that of 1000 indicates the average orbital period of the planet measured over the first 1000 years of  evolution. Using this measure, we aggregate the change in the periods of the planets from the initial conditions until a preset integration time is reached. The change in the period ratio between four-planet systems with well-spaced and closely-spaced initial longitudes can be seen in Figure \ref{fig:period shift}. Note that the closely-spaced Aligned longitudes are consistently shifted to a higher period ratio, but the shift diminishes as the initial separation grows. \textcolor{black}{This trend corresponds closely with the results in Table \ref{tab:4_planet_phase_shift} wherein the difference in period ratio for the phase shifting. This is exemplified by considering an initial separation of approximately $5.5, \, 7.0, \,\text{and } 8.0$ mutual Hill radii with a period shifting of $0.775 \times 10^{-3}, \, 0.548 \times 10^{-3}, \, \text{and } 0.284 \times 10^{-3}$.} For both SL09 and Aligned initial longitudes, there are significant increases in the period ratio that correspond to first-order resonances, especially between planets that are more closely-spaced. Additionally, small positive period shifts coincide with second-order MMRs between neighboring planets for the SL09 initial longitudes.

\section{Conclusions} \label{sec: conclusion}

We investigated the stability of four-planet systems of Earth-mass planets on closely-spaced, coplanar orbits around a solar mass star. The inner planet had an initial semimajor axis of 1 AU, and simulations were performed for $10^{10}$ virtual years or until the system reached instability, defined as a close encounter or planetary ejection. as further detailed in Section \ref{sec:methods}. An increase in initial planetary spacing resulted in a generally exponential increase in system lifetimes, similar to what was observed in previous work. However, proximity to mean motion resonances can reduce the system lifetimes by up to a few orders of magnitude, as observed in Figure \ref{fig:SL09Align}. Four-planet system lifetimes were investigated for multiple different combinations of initial longitudes and detailed in Section \ref{sec:results}. 

In aggregate, the lifetimes and general stability behavior of four-planet systems are more analogous to five-planet systems than three-planet systems 
(Section \ref{sec:345}). Similar to five-planet systems, four-planet systems lack the narrow regions of initial spacing with anomalously long-lived systems observed in three-planet systems by \cite{lissauer_gavino_2021}, as can be seen in Figure \ref{fig:3_4_5}. The geometrically-spaced three-, four-, and five-planet systems, however, still share regions of local minima in system lifetimes induced by MMRs. First-order resonances between neighboring planets have the strongest affects, but we also found reductions in system lifetimes associated first-order resonances between all pairs of non-neighboring planets, second-order resonances between neighboring planets and those non-neighboring planets with one planet between them, and even third-order resonances such as the 17/14 and 19/16 resonances between neighboring planets (Figures \ref{fig:SL09Align} and \ref{fig:third_order}).\\ 

An unpredicted result of comparing the lifetimes of systems in the Aligned and SL09 longitude sets is that there exists a progressive shifting of local maxima and minima in system lifetimes toward smaller initial semimajor axes separations for systems with initially Aligned longitudes compared to those with the widely-spaced SL09 longitudes. Uniformly and equally probable randomly-drawn initial longitudes, serving as a control, produced results close to those of SL09 and other sets of widely-spaced longitudes, indicating that the observed phase-shifting behavior is indeed induced by the specific choice of initial planetary alignments (Table \ref{tab:4_planet_phase_shift}).  In general, closely-spaced systems with initial longitudes at or approaching conjunction exhibit lifetimes more closely matching slightly more widely-spaced systems with well-separated initial longitudes (Fig.~\ref{fig:rollingmedians}).  However, the offset diminishes at wider orbital separations.  This shift is caused by the asymmetric energy transfers during the departure from the initial closely-aligned configuration. The effect is further detailed in Section \ref{subsec: Mechanism Responsible for Phase Shifting}. Figure \ref{fig:period shift} demonstrates this effect can also be observed through shifts in mean planetary orbital periods. \\

Due to the aforementioned phase-shifting behavior, for future studies, we recommend using either well-spaced or random longitudes, where in most cases the preferred method should be the latter. Both methodologies  mitigate the systematic phase shifting observed, assuming that nominal results are desired. Using random longitudes necessitates higher resolution to overcome the random scatter for longitudinal comparison, or duplicitous sampling for a given radial spacing. This results in greater computational time required for this method, where using well-spaced longitudes could be used as a lower-cost substitute.

\section*{Acknowledgments}
 BJO and GEN thank the United States Department of Defense Science, Math, and Research for Transformation (SMART) Scholarship for Service for educational funding. BJO dedicates his contribution {\footnotesize \calligra S.D.G}. JJL was supported in part through NASA's PSD ISFM program. We thank Pierre Gratia for providing a Python script for integrating with Rebound that was used as a guide for the script used for this work and Tjarda Boekholt and Jacob Kegerreis for helpful comments on draft versions of this manuscript. 

\bibliography{ref}

\appendix \label{sec:appendix}
\vspace{5mm}

\section{Third-order Resonances} \label{sec:third-order Resonances}

The top panel in Figure \ref{fig:SL09Align} shows that almost all four-planet systems with SL09 initial longitudes and $\beta \approx 9$ survive for $> 10^{10}$ years, with the single exception at 0.01 resolution within the range $8.91 \leq \beta \leq 9.21$ located at 9.08, very close to the 19/16 MMR between neighboring planets at $\approx$~9.0797. Here, we present results of integrations near the 19/16 using a finer grid in initial orbital separation of $\beta = 0.001$ over the range $9.07 \leq \beta \leq 9.09$. 

\begin{figure}[H]
    \centering
    \includegraphics[width=0.6\linewidth]{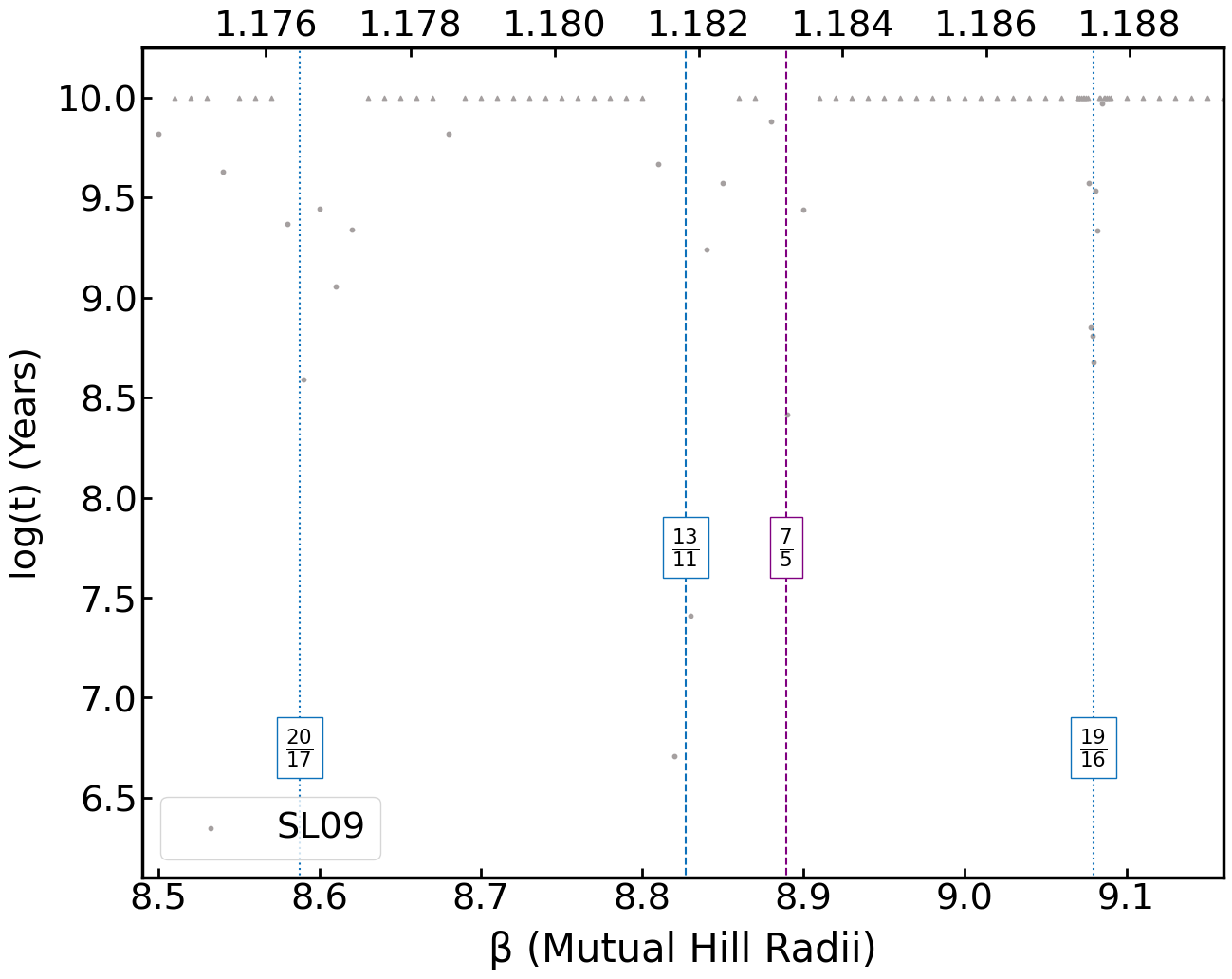}
    \caption{Close-up view of four-planet systems with SL09 longitudes over the interval $\beta \in [8.5, 9.15]$ and corresponding values in initial period ratio on the top axis. Three different types of MMRs are denoted: dashed blue is a second-order resonance between neighboring planets, dashed purple is a second-order resonance between planets with one planet between them, and the dotted blue lines are third-order resonances between neighboring planets. Near the $20/17$, $13/11$, and $7/5$ resonances, the resolution is 0.01 in mutual Hill radii. However, the $19/16$ resonance has a resolution of 0.001 in mutual Hill radii over the range $\beta \in [9.07, 9.09]$ to compensate for the narrowness of this feature. 
    }
    \label{fig:third_order}
\end{figure}

Figure \ref{fig:third_order} displays the results of these integrations as well as a zoomed-in version of the data shown previously covering the region from slightly inwards of the 20/17 MMR between neighboring planets to a bit beyond the 19/16 MMR. A sharp, narrow region of reduced lifetime centered on the 19/16 resonance is clearly evident with the 6 systems with spacings of 9.077~--~9.082 mutual Hill radii becoming unstable prior to the 10 gigayears termination time. The dip associated with the 20/17 is broader in $\beta$, but direct comparison between the effects of these two third-order resonances is problematic since only a lower bound is present for the baseline lifetimes of systems in the vicinity of each of these resonances and the dip associated with the 20/17 resonance appears to be centered slightly wide of exact resonance, suggesting that some other destabilizing affect may also be contributing to this feature.

\section{Phase Shift Quantification} \label{sec: Phase Shift Quantification}

\subsection{Calculating Locations of Extrema in System Lifetime} \label{subsec:Data Processing}
 
\textcolor{black}{To quantitatively locate the planetary spacings corresponding to the broad local maxima and minima in lifetimes observed, we sought a method to smooth the data sufficiently to mitigate scatter in lifetimes of systems with similar initial conditions caused by weak resonances, chance alignment, chaos, and other factors. We first applied the non-parametric Locally Weighted Scatterplot Smoothing (LOWESS) algorithm as formulated by \cite{cleveland} with a window of width $\beta$ = 0.1 to all investigated longitudes. This technique uses a local least-squares best-fit polynomial to smooth scattered data and does so piece-wise over a given window size. Our selected window size is sufficiently large to smooth out variations in system lifetimes occurring over narrow ranges of planetary separation and small enough to avoid smoothing over the valleys in lifetimes caused by first-order and second-order resonances and the peaks between them, as shown in Figure \ref{fig:rollingmedians}.  We compare data sets at resolutions of $\beta = 0.001$ for all of our initial longitude prescriptions with the exception of the Random longitudes.  The Random longitude data has a resolution of $\beta=0.0005$ to (at least partially) compensate for the extra scatter resulting from the addition of angular variations.}

 \textcolor{black}{After this first step of LOWESS smoothing, a technique prevalent in signal processing is utilized to determine the extrema locations. First, a search is performed to determine ``extrema candidates", regions wherein a sign change in the difference between adjacent lifetime values in the dataset occurs. Next, the extrema candidates are further filtered based upon topological prominence, the difference between the lifetimes of extrema and the relative local contour. We further restrict identified extrema candidates to peaks in lifetime separated by more than $0.1$ in $\beta$ and troughs separated likewise. Note, this is still maintained for T-4.6 and P-4.6 extrema, since the minimum separation requirement is for peak to peak and trough to trough. Finally, spurious extrema are removed by specifying a minimum width of an extremum of $\beta \ge 0.02$  and rejecting plateaus (adjacent peaks with no intervening valleys). Following the smoothing and filtering operation, the most prominent extrema are extracted and may be compared  between datasets. The remaining extrema are labeled in Figure \ref{fig:rollingmedians} and compared in Table \ref{tab:4_planet_phase_shift}.}

 \textcolor{black}{To estimate the uncertainty of this extrema detection method, we performed a cross-correlation test comparing the locations of the extrema calculated from two sets of systems with SL09 longitudes. One of these sets is our standard set of runs spanning $3.465 \leq \beta \leq 8.200$ at a resolution of 0.001; the comparison set of systems spans $3.4645 \leq \beta \leq 8.1995$ at the same resolution.  This dataset, referred to as ``SL09 Shift'' in Table \ref{tab:4_planet_phase_shift}, differs by having initial orbital separations offset in $\beta$ by 0.0005. The magnitude of the differences between prominent peaks and troughs in these data sets was used to compute an estimate of the method uncertainty. A t-distribution was utilized with a $3\sigma$ confidence found to be $\Delta\beta\approx 0.0042$ as an estimate of uncertainty in locations of extrema in differing sets of runs. }

\subsection{Quantification Algorithm}
\begin{figure}[H]
    \centering
    \includegraphics[width=0.95\linewidth]{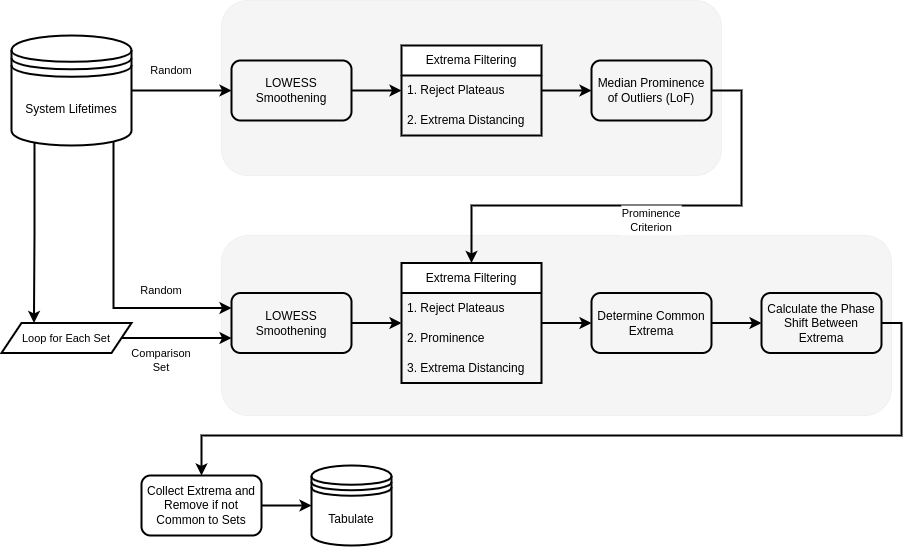}
    \caption{Process flow for the phase shift extraction algorithm. From the set of system lifetimes, different sets of longitudes are compared. The base set used as a datum are the Random longitudes and are processed in accordance with the top process in the diagram. Through this process, an unbiased criterion can be determined. The remaining lifetimes corresponding to the longitude sets are compared in the bottom process to determine the phase shift. Finally, the common phase shifts between each set of system lifetimes are extracted.}
    \label{fig:alg}
\end{figure}

For the quantification of the phase shift, a composite technique is employed to select the most distinct extrema from the set of candidate extrema; this process is summarized in Figure \ref{fig:alg}. Each component is discussed in greater detail below. Finding extrema in noisy data is a common exercise in the field of signals processing. However, additional constraints need to be placed upon extrema searches to only accept strongly present extrema for proper comparison. The weakness in classical approaches is due to the relatively strong effects of the local scatter in system lifetimes; to rectify this, we develop a new algorithm. 

\subsection{LOWESS Smoothing}
Robust Locally Weighted Regression and Smoothing Scatterplots (LOWESS), developed by \cite{cleveland}, is a robust optimal polynomial fitting technique for smoothing noisy data. This technique is useful because it results in a differentiable relationship from heavily scattered data while maintaining the local behavior. This is accomplished by weighting an aggregation of local polynomial fits in the region. There are also robustness constraints to reject outliers that may occur, in order to better represent the bulk trend.




\subsection{Extrema Detection}
Once the system lifetime dataset has been smoothed, a search for the local extrema can be employed. The underlying search is simply comparing the relative differences between data points until the difference undergoes a sign change. Even after smoothing, undesirable extrema will be detected. Therefore, some additional constraints must be applied to reduce the number of spurious extrema detected, e.g., plateau rejection, extrema distancing requirements, and extrema prominence.  Table \ref{tab:SmoothingParameters} gives relevant parameters used.  

\subsubsection{Plateau Rejection}
One of the features that is apparent after smoothing is the existence of local plateau `peaks', which are broad maxima whose peak location may be offset significantly due to the effect of chaotic and hyper-local scatter in the data. Hence, such plateaus should be rejected from the set of peaks.  To distinguish a strong extremum from a plateau, we require significant change in magnitude between the slopes of best fit lines of the local regions on each side of the extremum candidate (set as 0.03 mutual Hill radii). Two examples of plateaus that should be rejected are those between $5.5 \leq \beta \leq 6.0$ as well as $6.5 \leq \beta \leq 7.0$ in Figure \ref{fig:rollingmedians}.

\subsubsection{Extrema Distancing}
There may be a cluster of local extrema that are close together with some being stronger than the other local extrema. For such small local regions, we desire only one maximum or one minimum if the other criterion for peaks are met. Therefore, the strongest extremum in the local region is selected and the others are discarded. For the compact initial spacings and number of planets we are studying, the prominent extrema of interest will be correlated with a first or second order resonance or a "recovery" from a trough in system lifetime between mean-motion resonances. Therefore, throughout all datasets investigated, especially when smoothed, the extrema of interest will correspond to the same mean-motion resonances.

\subsubsection{Prominence Criterion}
Commonly used in determining mountain peaks, prominence is a measure of the minimum elevation one must lose when descending from a peak before beginning to climb another neighboring peak. In the 2-D representation of peaks and valleys, prominence is the difference in $\log {t_c}$ between a local maxima and the highest local minima on either side of it.  The converse is true for valleys, where we consider prominence as the $\log {t_c}$ difference between the valley and lowest neighboring peak. This prominence criterion is useful for rejecting small local fluctuations before or after the true extrema value. 

However, determining a sufficient threshold for prominence is difficult. Confirmation bias should be minimized to avoid picking a prominence value to select certain peaks instead of others. To rectify this, following Figure \ref{fig:alg}, all of the criterion are applied without a prominence criterion. The most prominent extrema are extracted by determining the outliers of the extrema in terms of prominence using a technique called Local Outlier Factor formulated by \cite{10.1145/335191.335388}. 
The median of these outlier extrema prominence are taken and used as criterion for the prominence. 

\subsection{Post-Processing}
After the extrema are determined, the phase shift between two system lifetime datasets is determined. Given the set of extrema between the two datasets, the extrema are collated into a single set by selecting the extrema that appear in both extrema sets. Ultimately, there is a creation of a set of extrema that are common to both sets. Afterward, the locations of common extrema for each dataset are subtracted from the locations of the extrema in the Random data set to determine the phase shift.

\subsection{Table of Constants Used}

\begin{table}[h!]
    \centering
    \tiny
    \begin{tabular}{c|c|c}
        Constant & Value & Reason \\
        \hline
        \hline
        LOWESS Fraction & 0.1 mutual Hill radii & Follows the value from below  \\
        Distance Between Like-Extrema & 0.1 mutual Hill radii & Significantly smaller than the closest noticeable extrema \\
        Points on each side for Plateau Rejection & 0.03 mutual Hill radii & Robustness limit for maintaining consistency with changes in dataset \\
        Slope Criterion & 3.0 & Rough classification consistency limit \\
        \end{tabular}
    \caption{Important constants used in determining the location of extrema. The values regarding the plateau rejection were roughly generated via robustness tests between a variety of system lifetime datasets to ensure that plateaus were properly rejected. The authors stress that these constants are somewhat arbitrary and were partially determined qualitatively.}
    \label{tab:SmoothingParameters}
\end{table}








\end{document}